\documentstyle[myart,amsfonts,12pt]{article}

\normalbaselines
\font\tenbm=cmmib10
\font\sevenbm=cmmib7
\font\fivebm=cmmib5
\newfam\bmfam
\textfont\bmfam=\tenbm \scriptfont\bmfam=\sevenbm
\scriptscriptfont\bmfam=\fivebm
{\count0=\number\bmfam \multiply\count0 by "100
\def\defbgreek#1#2#3{{\count1=\count0 \advance\count1 by "#2#3
  \global\mathchardef#1=\count1 }}
\defbgreek\balpha  0B \defbgreek\brho       1A
\defbgreek\bbeta   0C \defbgreek\bsigma     1B
\defbgreek\bgamma  0D \defbgreek\btau       1C
\defbgreek\bdelta  0E \defbgreek\bupsilon   1D
\defbgreek\bepsilon0F \defbgreek\bphi       1E
\defbgreek\bzeta   10 \defbgreek\bchi       1F
\defbgreek\bmeta   11 \defbgreek\bpsi       20
\defbgreek\btheta  12 \defbgreek\bomega     21
\defbgreek\biota   13 \defbgreek\bvarepsilon22
\defbgreek\bkappa  14 \defbgreek\bvartheta  23
\defbgreek\blambda 15 \defbgreek\bvarpi     24
\defbgreek\bmu     16 \defbgreek\bvarrho    25
\defbgreek\bnu     17 \defbgreek\bvarsigma  26
        \defbgreek\bxi     18 \defbgreek\bvarphi    27
\defbgreek\bpi     19}
\input tcilatex

\begin{document}

\title{Pair Production Problem and Canonical Quantization of Nonlinear 
Scalar Field in Terms of World Lines}
\author{Yuri A.Rylov}
\date{Institute for Problems in Mechanics, Russian Academy of Science,\\
101 Vernadskii Ave., Moscow, 117526, Russia.}
\maketitle

\begin{abstract}
A new quantization scheme ($WL$-scheme), using world lines as objects of
quantization, is proposed. Applying to nonlinear scalar field, the $WL$%
-scheme is investigated and compared with the conventional $PA$-scheme of
quantization. In the $PA$-scheme objects of quantization are particles and
antiparticles, which are fragments of the total physical object -- world
line (WL). Applying to the nonlinear field, the $PA$-scheme of quantization
leads to such difficulties as nonstationary vacuum, obligatory use of
perturbation theory technique, normal ordering and cut-off at $t\rightarrow
\infty $ in the scattering problem. These difficulties are corollaries of
inconsistency of $PA$-scheme. The $WL$-scheme is free of these difficulties.
These difficulties are connected with the reconstruction problem of the
total world lines from their fragments (particles and antiparticles). In the
case, when these fragments interact between themselves, such a
reconstruction is very complicated problem. The new $WL$-scheme of
quantization is free of all these problems, because it does not cut the
total world line into fragments (particles and antiparticles). Formally
appearance of fragments in the conventional quantization $PA$-scheme is a
corollary of identification of the energy with the Hamiltonian. In fact such
an identification is not necessary. It leads only to difficulties. The new $%
WL$-scheme of quantization does not use this identification and enables to
go around all these problems. The $WL$-scheme enables not to use additional
(to nonrelativistic QM) quantization rules, used in the relativistic QFT
(normal ordering, perturbation technique, renormalization).
\end{abstract}

\newpage

\section{Introduction}

The problem of pair production is the crucial problem of relativistic
quantum field theory (QFT), as well as that of the elementary particles
theory. In this paper one investigates, if it is possible to describe the
pair production in the scope of conventional quantum mechanics principles,
i.e. without using additional quantum rules of QFT such as normal ordering,
perturbation theory technique, manipulations with nonstationary vacuum,
renormalization and interaction cut-off in the scattering problem. The main
result of the investigation is formulated as follows. The secondary
quantization of the nonlinear relativistic scalar field with the self-action
term $\lambda \varphi ^{\ast 2}\varphi ^{2}$ and its description without the
perturbation theory is possible. The vacuum state appears to be stationary
and the normal ordering is not used. In other words, one succeeds to
overcome many problems of relativistic QFT, but the problem of pair
production remains to be unsolved. This problem appears to be more subtle
and complicated, than it is common practice to think\footnote{%
This investigation has long been carried out (before 1990), but we failed to
publish it, because the editors of physical journals believed that one may
not publish the paper, where the secondary qunatization of nonlinear field
do not produce pair production, even if the procedure of secondary
quantization is completely consistent. Nevertheless results of this
investigation were a motivation for a search of an alternative approach to
the pair production problem. Now, when alternative description of pair
production has appeared \cite{R98}, publication of this investigation may be
interesting as an argument in favour of this alternative.}.

The fact is that the secondary quantization of a relativistic field is
accompanied usually by fragmentation of world lines into particles and
antiparticles. Some fragments of the particle world line (WL) describe
particles, other one describe antiparticles. Appearance of the perturbation
theory in QFT is essentially a result of this fragmentation. Description of
the fragmentation process is rather simple, whereas description of
reciprocal process of defragmentation is more complicated and description of
defragmentation is imperfect. In some cases the pair production arises as a
result of the defragmentation process, but not as a result of turn of the
world line in time.

Application of the world line fragmentation (perturbation theory technique)
in QFT is connected closely with identification of Hamiltonian $p_{0}\equiv
-H$ with the energy $E$. The energy is defined by the relation
\begin{equation}
E=P^{0}=\int T^{00}d{\bf x}  \label{b1.1}
\end{equation}
where $T^{ik}$ is the energy-momentum tensor, whereas $p_{0}\equiv -H$ is
defined as the quantity canonically conjugate to time. In general, the
energy $E$ and Hamiltonian $H$ are different quantities, but in the
nonrelativistic case they coincide for free particle, i.e. $E=-H,$ and it is
common practice to think that this relation takes place in QFT, where it has
the form of the relation
\begin{equation}
\lbrack u,P_{k}]_{-}=-i\hbar \frac{\partial u}{\partial x^{k}},
\label{a1.19}
\end{equation}
Here $P_{k}$ is the energy-momentum operator, $[u,P_{k}]_{-}$ is commutator,
and $u$ is operator of arbitrary dynamic variable. $\hbar $ is the Planck
constant which is set to be equal to 1 further. It is common practice to
think \cite{BS59} that the relation (\ref{a1.19}) is necessary for
determination of relativistic commutation relations. Use of the relation (%
\ref{a1.19}) is the ground for the statement that the field operator $%
\varphi $ cannot contain only annihilation operators (as in nonrelativistic
case) and that $\varphi $ contains both creation and annihilation operators.
The last generates a necessity of a use of the perturbation theory methods
in relativistic QFT.

Eliminating relation (\ref{a1.19}), one can use such a secondary
quantization of relativistic field, where operator $\varphi $ contains only
annihilation operators, and Hermitian conjugate operator $\varphi ^{\ast }$
contains only creation operators. Commutation relations for operators $%
\varphi $ and $\varphi ^{\ast }$ can be determined without a reference to
relation (\ref{a1.19}). They are determined from their relativistic
covariance and dynamic equation for $\varphi $ \cite{R72}.

It was shown \cite{R70} that the relation $E=-H,$ (or $E=H$) takes place in
the relativistic case only if particles and antiparticles are considered to
be independent physical objects, (but not different states of a physical
object WL, described by world line). Thus, the identification of $E$ and $H$
(relation $E=-H$) means fragmentation of WL into fragments, describing
independent physical objects particles and antiparticles. The idea of using
the world line as a physical object is the old idea. It goes back to
Stueckelberg \cite{S42} and Feynman \cite{F49}. Unfortunately, it was
developed somewhat inconsistently, and origin of this inconsistency is the
relation (\ref{a1.19}). In general, the energy $E$ and the Hamiltonian $H$
are defined independently, and there is no reason for their forced
identification. If they coincide for some reason (for instance, in force of
dynamic equation), they will coincide independently of the relation (\ref
{a1.19}). If they do not coincide, there is no reason for making them to be
equal.

Farther we shall use and compare two schemes of canonical quantization.
Conventional quantization scheme, using relation (\ref{a1.19}), will be
referred to as $PA$-scheme. The quantization scheme which does not use the
relation (\ref{a1.19}) and considers the world line as a physical object
will be referred to as $WL$-scheme. In general, the special term ''WL'' will
be used for the world line, considered to be a physical object.

Before consideration of different modifications of canonical quantization
one should understand distinction between $WL$-description and $PA$%
-description on the classical level.

Let $x^{i}$ be coordinates of a point of the space-time in some coordinate
system, and
\begin{equation}
{\cal L}:\qquad x^{i}=q^{i}(\tau ),\qquad i=0,1,2,3  \label{a1.1}
\end{equation}
be a continuous world line, $\tau $ being a parameter along it. Let ${\cal L}
$ describe a history of particles and antiparticles produced under influence
of some given external field $f(q)$ which can produce pairs. WL ${\cal L}$
is described by the following parameters: the mass $m$, and the ''charge'' $%
e $. The mass is non-negative constant which describes interaction with the
gravitational field. The ''charge'' $e$ is a constant, describing
interaction with the electromagnetic field. The orientation ${\bf %
\varepsilon }$ is one more Lorentz invariant quantity, describing WL. It is
a discrete dynamic variable, describing the state of WL. Changing the
orientation ${\bf \varepsilon }$ (with fixed $m$ and $e$), one turns a
particle to an antiparticle and vice versa. Orientation ${\bf \varepsilon }$
describes one of two possible directions of motion along the world line. If
the parametrization ${\cal P}$ of ${\cal L}$ is realized by the parameter $%
\tau $, then the orientation ${\bf \varepsilon }$ is determined by the
component $\varepsilon $ at this parametrization ${\cal P}$. The component $%
\varepsilon $ takes values $\pm 1.$ At transformation of the WL ${\cal L}$
parametrization
\begin{equation}
{\cal P}\rightarrow {\cal P}^{\prime },\qquad \tau \rightarrow \tau ^{\prime
}=\tau ^{\prime }(\tau ),\qquad \partial \tau ^{\prime }/\partial \tau
\not\equiv 0  \label{a1.2}
\end{equation}
the orientation component transforms according to the law
\begin{equation}
\varepsilon \rightarrow \varepsilon ^{\prime }=\varepsilon \hbox{ sgn}{\frac{%
\partial \tau ^{\prime }(\tau )}{\partial \tau }}  \label{a1.3}
\end{equation}
For WL ${\cal L}$ the action has the form
\begin{equation}
S[q]=\int\limits_{\min (\tau ^{\prime },\tau ^{\prime \prime })}^{\max (\tau
^{\prime },\tau ^{\prime \prime })}Ld\tau ,  \label{a1.4}
\end{equation}
\begin{equation}
L=-\sqrt{m^{2}c^{2}\dot{q}^{i}g_{ik}\dot{q}^{k}-\alpha f(q)}-\frac{%
\varepsilon e}{c}\dot{q}^{i}A_{i}(q),\qquad \dot{q}^{i}\equiv dq^{i}/d\tau ,
\label{b1.5}
\end{equation}
where $\tau ^{\prime }$ and $\tau ^{\prime \prime }$ are values of the
parameter $\tau $ at the integration interval boundaries. $\alpha $ is a
non-negative constant. Here $f(q)$ is a given external field which can turn
the world line in the time direction, i.e. it can create or annihilate
particle-antiparticle pairs. The fact is that the Lagrangian (\ref{a1.4})
with $\alpha =0$ admits only timelike WLs (\ref{a1.1}), ($\dot{q}^{i}\dot{q}%
_{i}>0$ takes place everywhere). Introduction of the term $\alpha f(q)$
removes this constraint. This capacity of pair production remains in the
limit $\alpha \rightarrow +0,$ when the action (\ref{a1.4}) becomes to be
invariant with respect to arbitrary transformation (\ref{a1.2}) of
parametrization (see details in ref. \cite{R70}). The radical in Eq.(\ref
{a1.4}) is supposed to be non-negative. The action (\ref{a1.4}), (\ref{b1.5}%
) is invariant with respect to arbitrary coordinate transformation. In the
limit $\alpha \rightarrow +0,$ it becomes to be invariant also with respect
to arbitrary parametrization transformation (\ref{a1.2}).

The electric current $J^{i}(x)$ and energy-momentum tensor $T^{ik}$ are
defined as sources of electromagnetic field and gravitational one
respectively. In the Galilean coordinate system, where $g_{ik}$= const and $%
A_{i}=0,$ they have the form
\begin{equation}
J^{i}(x)=-c\frac{\delta S}{\delta A_{i}(x)}=\varepsilon e\sum_{l}\frac{\dot{q%
}^{i}(\tau _{l})}{\left| \dot{q}^{0}(\tau _{l})\right| }\delta ({\bf x}-{\bf %
q}(\tau )),\qquad i=0,1,2,3  \label{a1.5}
\end{equation}
\begin{equation}
T^{ik}(x)=-{\frac{2c}{\sqrt{\left| g\right| }}}{\frac{\delta S}{\delta g_{ik}%
}}=\sum_{l}\frac{mc^{2}\dot{q}^{i}(\tau _{l})\dot{q}^{k}(\tau _{l})}{\sqrt{%
\dot{q}^{l}(\tau _{l})g_{lj}\dot{q}^{j}(\tau _{j})}}\frac{\delta ({\bf x}-%
{\bf q}(\tau _{l}))}{\left| \dot{q}^{0}(\tau _{l})\right| }  \label{a1.6}
\end{equation}
\[
i,k=0,1,2,3
\]
where $\tau _{l}=\tau _{l}\left( x^{0}\right) $ are roots of the equation
\begin{equation}
q^{0}(\tau _{l})-x^{0}=0  \label{a1.7}
\end{equation}
and
\[
g=\det \left| \left| g_{ik}\right| \right| ,\qquad i,k=0,1,...n.
\]
Intersection of the WL with the plane $x^{0}$= const consists of one or
several points. The following quantities can be attributed to such a point:
canonical momentum $p_{i}$ $(i=0,1,2,3)$, electric charge $Q$, energy $E$,
and momentum $P_{\beta }$ $(\beta =1,2,3)$. They are defined as follows.
\begin{equation}
p_{i}=\frac{\partial L}{\partial \dot{q}^{i}}=-\frac{mc\dot{q}_{i}}{\sqrt{%
\dot{q}^{l}g_{lk}\dot{q}^{k}}},\qquad i=0,1,2,3  \label{a1.8}
\end{equation}
\begin{equation}
Q=\int J^{0}(x)d{\bf x}=\varepsilon e\hbox{ sgn}(\dot{q}^{0})=-\varepsilon e%
\hbox{ sgn}(p_{0})  \label{a1.9}
\end{equation}
\begin{equation}
P^{0}=E=\int T^{00}d{\bf x}=\frac{mc\left| \dot{q}^{0}\right| }{\sqrt{\dot{q}%
^{l}g_{lk}\dot{q}^{k}}}=\left| p_{0}\right|  \label{a1.10}
\end{equation}
\begin{equation}
P^{\alpha }=\int T^{0\alpha }d{\bf x}=\frac{mc\dot{q}^{\alpha }}{\sqrt{\dot{q%
}^{l}g_{lk}\dot{q}^{k}}}\hbox{sgn}(\dot{q}^{0})=-p^{\alpha }\hbox{sgn}%
(p_{0}),\qquad \alpha =1,2,3  \label{a1.11}
\end{equation}
One can attribute two invariants: $m=\sqrt{p_{i}p^{i}}$,\ \ $Q=\pm e$ to any
point of intersection of ${\cal L}$ with $x^{0}$= const. These points will
be referred to as SWLs. (Abbreviation (SWL) of the term ''section of world
line''. SWLs can be of two kinds: $(m,e)$ and $(m,-e)$. One of them is a
particle, and other one is an antiparticle. Thus, a SWL is a collective
concept with respect to concepts of a particle and an antiparticle.

The SWL state is described either by dynamic variables
\begin{equation}
x^{\alpha },\varepsilon _{p},p_{\alpha },\qquad \alpha =1,2,3,  \label{a1.12}
\end{equation}
or by observable physical variables
\begin{equation}
x^{\alpha },Q,P^{\alpha },\qquad \alpha =1,2,3  \label{a1.13}
\end{equation}
where $\varepsilon _{p}$ takes values $\pm 1$, which label particle and
antiparticle. Dynamic variables $x^{\alpha },\varepsilon _{p},p_{\alpha
},\quad \alpha =1,2,3$ are connected with physical ones by relations

\begin{equation}
\varepsilon _{p}=\hbox{ sgn}(p_{0}),\qquad p_{0}=\varepsilon _{p}E,\qquad
P_{\alpha }=-\varepsilon _{p}p_{\alpha },  \label{a1.14}
\end{equation}
\[
Q=-\varepsilon \varepsilon _{p}e,\qquad E=\left| \sqrt{{\bf p}^{2}+m^{2}}%
\right| ,
\]

Let us consider such transformations of the way of the WL description which
do not change conservative quantitites $Q,E,P_{\alpha }$.
\[
I_{\tau }:\qquad \tau \rightarrow -\tau ,\qquad x^{i}\rightarrow
x^{i},\qquad \varepsilon \rightarrow -\varepsilon ,\qquad \varepsilon
_{p}\rightarrow -\varepsilon _{p},
\]
\begin{equation}
p_{\alpha }\rightarrow -p_{\alpha },\qquad Q\rightarrow Q,\qquad P_{\alpha
}\rightarrow P_{\alpha },  \label{a1.15}
\end{equation}
\[
I_{ey}:\qquad \tau \rightarrow \tau ,\qquad x^{i}\rightarrow
-(x^{i}-2y^{i}),\qquad e\rightarrow -e,\qquad \varepsilon _{p}\rightarrow
-\varepsilon _{p},\qquad \varepsilon \rightarrow \varepsilon ,
\]
\begin{equation}
\qquad p_{\alpha }\rightarrow -p_{\alpha },\qquad Q\rightarrow Q,\qquad
P_{\alpha }\rightarrow P_{\alpha },  \label{a1.16}
\end{equation}
where $y=y^{i}$ is a transformation parameter. Both transformations $I_{\tau
}$ and $I_{ey}$ change the sign of the canonical momentum $p_{i}$ but do not
change the energy-momentum vector $P_{i}$ and the charge $Q$. Transformation
$I_{\tau }$ changes parametrization of WL (parameter $\tau $), but does not
change parameters $\left( m,e\right) $ of dynamic system WL. Vice versa,
transformation (\ref{a1.16}) does not change parametrization of WL, but
changes parameters $\left( m,e\right) \rightarrow \left( m,-e\right) $ of
dynamic system WL, described by the action (\ref{a1.4}), (\ref{b1.5}) and
its state $\left( x,\varepsilon _{p},p_{\alpha }\right) \rightarrow \left(
-x+2y,-\varepsilon _{p},-p_{\alpha }\right) $.

Two WLs $\left( m,e\right) $ and $\left( m,-e\right) $ are two different
dynamic systems, whereas $\left( x,\varepsilon _{p},p_{\alpha }\right) $ and
$\left( x,-\varepsilon _{p},-p_{\alpha }\right) $ with the same $\left(
m,e\right) $ are two different states of the same dynamic system WL. In
other words, transformation (\ref{a1.15}) does not change the dynamic system
WL, described by the action (\ref{a1.4}), (\ref{b1.5}), whereas the
transformation (\ref{a1.16}) does change dynamic system WL in itself.

When there is no pair production and WL does not make zigzag in the time
direction, one can achieve coincidence of vectors $p_{i}$ and $-P_{i}$. In
this case the energy-momentum vector $-P_{i}$ is canonically conjugate to
the vector $x^{i}$.

If WL describes the pair production (see Fig.1), the coincidence of $p_{i}$
and $-P_{i}$ can be achieved by means of transformation (\ref{a1.15})
performed only on those WL interval, where $p_{i}$ does not coincide with $%
-P_{i}$. For instance, one can choose $\tau = x^{0}$, then $p_{i} = -P_{i}$
everywhere, but the parameter $\tau$ will change non-monotonically along the
WL.

The description, where $P_{i}=-p_{i}$, will be called $PA$-description. The
approach, using such a non-monotone parametrization, will be referred to as
the $PA$-approach (the approach from the standpoint of particles and
antiparticles). The approach, where a monotone parametrization of WLs is
used, will be referred to as $WL$-approach. Thus, the $WL$-approach
distinguishes between the canonical momentum and energy-momentum, the $PA$%
-approach does not. A criterion of the $PA$-approach is the condition.
\begin{equation}
p_{i}=-P_{i},\qquad i=0,1,2,3,  \label{a1.18}
\end{equation}
that the energy-momentum vector $-P_{i}$ defined by Eqs. (\ref{a1.10}), (\ref
{a1.11}) were canonically conjugate to $x^{i}$.

From the standpoint of $WL$-approach the non-monotone parametrization of the
WL is not consistent, as far as such non-monotony is absent in
non-relativistic description. Vice versa, from the standpoint of $PA$%
-approach the separation of concepts of the energy-momentum $P_{i}$ and the
canonical momentum $p_{i}$ is not satisfactory, because such a separation is
absent in non-relativistic mechanics.

From the standpoint of Stueckelberg-Feynman idea that the world line is a
physical object, $WL$-approach is more consistent, than $PA$-approach.

Distinction between the two approaches can be manifested only at pair
production, i.e. only in the quantum field theory (QFT). The canonical
quantization in QFT uses $PA$-approach, i.e. Eq.(\ref{a1.18}) is fulfilled
always.

In QFT the condition (\ref{a1.18}) takes the form (\ref{a1.19}). From the
non-relativistic viewpoint the $PA$-approach is more clear, than $WL$%
-approach. But a use of $PA$-approach at a quantization of interacting
fields leads to a set of difficulties. The principal difficulty is absence
of a stationary vacuum state.

In the present paper one investigates application of the $WL$-approach to
the Boson field quantization. In Sec.2 the equivalency of $WL$-approach and $%
PA$-approach at the free field quantization is shown. In section 3 the
concept of quantization model is introduced. Quantization of nonlinear field
is investigated in Section 4. Section 5 is devoted to description of
nonlinear field in terms of free fields. In section 6 the scattering problem
without interaction cut-off at $t\rightarrow \infty $ is investigated. In
section 7 some peculiarities of introducing physical quantities for
nonlinear field are considered.

\section{Quantization of the Free Scalar Field}

Let us consider a charged scalar field $\varphi $ in the $(n+1)$-dimensional
space-time. $x^{i}=(x^{0},{\bf x})=(x^{0},x^{\alpha })$ are Galilean
coordinates. Latin indices take values $0,1,\ldots n$, Greek ones do $%
1,2,\ldots n$. As usually a summation is made on like super- and subscripts.
Let the operator $\varphi (x)=\varphi ({\bf x},t)$ satisfy the equation
\begin{equation}
(\partial _{i}\partial ^{i}+m^{2})\varphi =0  \label{a2.1}
\end{equation}
$\varphi ^{\ast }(x)$ be a corresponding Hermitian conjugate operator.

It is convenient to use variables
\begin{equation}
b(\varepsilon _{k},{\bf k},t)=b(K,t)=i{(2\pi )}^{-n/2}\int\limits_{-\infty
}^{\infty }\exp (-i{\bf kx})\left[ {\frac{\sqrt{\beta ({\bf k})}}{2}}\varphi
({\bf x},t)+\frac{i\varepsilon _{k}}{\sqrt{\beta ({\bf k})}}\dot{\varphi}(%
{\bf x},t)]\right] d{\bf x}  \label{a2.2}
\end{equation}
\[
K=\{\varepsilon _{k},{\bf k\}},\qquad \dot{\varphi}\equiv \partial \varphi
/\partial t,\qquad d{\bf x\equiv }dx^{1}dx^{2}\ldots dx^{n},
\]
where $\varepsilon _{k}$=sgn$(k_{0})$ describes orientation.
\[
k_{0}=\varepsilon _{k}E({\bf k}),\qquad E({\bf k})=\left| \sqrt{m^{2}+{\bf k}%
^{2}}\right| ,\qquad \beta ({\bf k})=2E({\bf k}),
\]
\begin{equation}
{\bf kx}=-k_{\alpha }x^{\alpha }=k^{\alpha }x^{\alpha },\qquad k=(k_{0},{\bf %
k})  \label{a2.3}
\end{equation}
The reciprocal transformation has the form
\begin{equation}
\varphi ({\bf x},t)={(2\pi )}^{-n/2}\int\limits_{-\infty }^{\infty }e^{i{\bf %
kx}}\frac{b(K,t)}{\sqrt{\beta ({\bf k})}}dK  \label{a2.4}
\end{equation}
\begin{equation}
\dot{\varphi}({\bf x},t)=\frac{\partial \varphi }{\partial t}=-i{(2\pi )}%
^{-n/2}\int\limits_{-\infty }^{\infty }{\frac{\varepsilon _{k}}{2}}\sqrt{%
\beta ({\bf k})}e^{i{\bf kx}}b(K,t)dK  \label{a2.5}
\end{equation}
where
\begin{equation}
\int (.)dK\equiv \sum_{\varepsilon _{k}=\pm 1}\int (.)d{\bf k},\qquad d{\bf k%
}=dk_{1}dk_{2}\ldots dk_{n},  \label{a2.6}
\end{equation}
The dynamic equation (\ref{a2.1}) in terms of $b(K,t)$ takes the form
\begin{equation}
\dot{b}(K,t)=-i\varepsilon _{k}E({\bf k})b(K,t)  \label{a2.7}
\end{equation}
or
\begin{equation}
b(K,t)=e^{-i\varepsilon _{k}E({\bf k})t}b(K),\qquad b(K)=b(K,0)  \label{a2.8}
\end{equation}
Connection between $\varphi ^{\ast }({\bf x},t),\dot{\varphi}^{\ast }({\bf x}%
,t)$ and $b^{\ast }(K,t)$ is obtained as a result of Hermitian conjugation
of Eqs.(\ref{a2.2}), (\ref{a2.4}), (\ref{a2.5}), (\ref{a2.7}), (\ref{a2.8}).

{\it Definition 2.1}. Quantization scheme is a totality of three relations:
(1) dynamic equation, (2) definition of a vacuum vector, (3) commutation
relation between the dynamic variable operators.

The Fock's representation is supposed to be used. In this representation
there is only one vacuum vector $|0\rangle $. Any state $\Phi $ can be
obtained as a result of acting a dynamic variable operator upon vacuum
vector $|0\rangle $.

The scheme of quantization in terms of particles and antiparticles $(PA$%
-scheme) is defined as follows
\begin{equation}
1:\qquad \dot{c}({\bf k},t)=-iE({\bf k})c({\bf k},t),\qquad \dot{d}({\bf k}%
,t)=-iE({\bf k})d({\bf k},t)  \label{a2.9}
\end{equation}
\begin{equation}
2:\quad
\begin{array}{lll}
c({\bf k})|0\rangle _{{\rm PA}}=0,\quad & d({\bf k})|0\rangle _{{\rm PA}%
}=0,\quad & c({\bf k})=c({\bf k},0) \\
_{{\rm PA}}\langle 0|c^{\ast }({\bf k})=0,\quad & _{{\rm PA}}\langle
0|d^{\ast }({\bf k})=0,\quad & d({\bf k})=d({\bf k},0)
\end{array}
\label{a2.10}
\end{equation}
\begin{equation}
\begin{array}{ll}
3a:\qquad & {[c({\bf k}),c^{\ast }({\bf k^{\prime }})]}_{-}={[d({\bf k}%
),d^{\ast }({\bf k^{\prime }})]}_{-}=\delta ({\bf k}-{\bf k^{\prime }}) \\
3b:\qquad & {[c({\bf k}),c({\bf k^{\prime }})]}_{-}={[d({\bf k}),d({\bf %
k^{\prime }})]}_{-}={[c({\bf k}),d({\bf k^{\prime }})]}_{-}=0
\end{array}
\label{a2.11}
\end{equation}
where
\begin{equation}
c({\bf k},t)=b(1,{\bf k},t),\qquad d({\bf k},t)=b^{\ast }(-1,{\bf k},t)
\label{a2.12}
\end{equation}
and $|0\rangle _{{\rm PA}}$ is the vacuum state vector.

The quantization scheme in terms of WLs ($WL$-scheme) has the form
\begin{eqnarray}
1 &:&\qquad \dot{b}(K,t)=-i\varepsilon _{k}E({\bf k})b(K,t)  \label{a2.13} \\
2 &:&\qquad b(K)|0\rangle _{{\rm WL}}=0,\qquad _{{\rm WL}}\langle 0|b^{\ast
}(K)=0  \label{a2.14} \\
&&
\begin{array}{ll}
3a:\qquad & [b(K),b^{\ast }(K^{\prime })]_{-}=\delta (K-K^{\prime })\equiv
\delta _{\varepsilon _{k},\varepsilon _{k^{\prime }}}\delta ({\bf k}-{\bf %
k^{\prime }}) \\
3b:\qquad & [b(K),b(K^{\prime })]_{-}=0
\end{array}
\label{a2.15}
\end{eqnarray}

Let us make a change of variables
\begin{equation}
c({\bf k},t)=b_{{\rm E}}(1,{\bf k},t),\qquad d({\bf k},t)=b_{{\rm E}}(-1,%
{\bf k},-t).  \label{a2.17}
\end{equation}
Substitution of relations (\ref{a2.17}) into Eqs. (\ref{a2.9}) -- (\ref
{a2.11}) leads to relations
\begin{eqnarray}
1 &:&\qquad \dot{b}_{{\rm E}}(K,t)=-i\varepsilon _{k}E({\bf k})b_{{\rm E}%
}(K,t)  \nonumber \\
2 &:&\qquad b_{{\rm E}}(K)|0\rangle _{{\rm PA}}=0,\qquad _{{\rm PA}}\langle
0|b_{{\rm E}}^{\ast }(K)=0  \label{a2.18} \\
3a &:&\qquad \lbrack b_{{\rm E}}(K),b_{{\rm E}}^{\ast }(K^{\prime
})]_{-}=\delta (K-K^{\prime })\equiv \delta _{\varepsilon _{k},\varepsilon
_{k^{\prime }}}\delta ({\bf k}-{\bf k^{\prime }})  \nonumber \\
3b &:&\qquad {\lbrack b_{{\rm E}}(K),b_{{\rm E}}(K^{\prime })]}_{-}=0
\nonumber
\end{eqnarray}

Comparing Eq. (\ref{a2.18}) with Eqs. (\ref{a2.13}) -- (\ref{a2.15}), one
can see that the $PA$-scheme is distinguished from the $WL$-scheme only with
designations. They are equivalent. The change of variables (\ref{a2.17})
which transforms $PA$-scheme into $WL$-scheme is not unique. There are as
many such transformations as there are Galilean coordinate systems.

Let us construct the field $\varphi _{{\rm E}}({\bf x},t)$, expressing it
through variables $b_{{\rm E}}(K,t)$ by means of Eqs.(\ref{a2.4}), (\ref
{a2.5}).
\begin{equation}
\varphi _{{\rm E}}({\bf x},t)=\frac 1{(2\pi )^{n/2}}\int e^{i{\bf kx}}\frac{%
b_{{\rm E}}(K,t)}{\sqrt{\beta ({\bf k})}}d{\bf k}  \label{a2.19}
\end{equation}

Using Eqs.(\ref{a2.12}), (\ref{a2.17}), (\ref{a2.2}), (\ref{a2.19}), one
obtains
\begin{equation}
\varphi _{{\rm E}}(x)=\hat{\Pi}_{+}\varphi (x)+\hat{\Pi}_{-}\hat{I}%
_{0}\varphi ^{\ast }(x),\qquad \hat{I}_{0}\equiv \left. \hat{I}_{y}\right|
_{y=0}  \label{a2.20}
\end{equation}
where $\hat{\Pi}_{+},\hat{\Pi}_{+},\hat{I}_{y}$ are operators acting on
functions of $x$.
\begin{equation}
\hat{\Pi}_{+}={\frac{1}{2\hat{E}}}(\hat{E}+i\hat{\partial}_{0}),\qquad \hat{%
\Pi}_{-}=\frac{1}{2\hat{E}}(\hat{E}-i\hat{\partial}_{0}),\qquad \hat{E}%
\equiv |(m^{2}+\partial _{\alpha }\partial ^{\alpha })^{1/2}|  \label{a2.21}
\end{equation}
\[
\hat{E}\varphi (t,{\bf x})=\frac{1}{(2\pi )^{2n}}\int \int e^{i{\bf k}({\bf x%
}-{\bf x}^{\prime })}E({\bf k})\varphi (t,{\bf x}^{\prime })d{\bf k}d{\bf x}%
^{\prime }
\]
and $\hat{I}_{y}$ is an operator of the coordinates $x^{i}$ reflection with
respect to the point $y^{i}$.
\begin{equation}
\hat{I}_{y}\varphi (x)=\varphi (2y-x)  \label{a2.22}
\end{equation}

If the function $\varphi (x)$ satisfies Eq. (\ref{a2.1}) then operators $%
\hat{\Pi}_{+}$ and $\hat{\Pi}_{-}$ are projection operators, having the
properties
\[
\hat{\Pi}_{+}\hat{\Pi}_{+}\varphi =\hat{\Pi}_{+}\varphi ,\qquad \hat{\Pi}_{-}%
\hat{\Pi}_{-}\varphi =\hat{\Pi}_{-}\varphi ,\qquad \hat{\Pi}_{+}\hat{\Pi}%
_{-}\varphi =\hat{\Pi}_{-}\hat{\Pi}_{+}\varphi =0,
\]
\begin{equation}
(\hat{\Pi}_{+}+\hat{\Pi}_{-})\varphi =\varphi ,\qquad \hat{\Pi}_{+}\hat{I}%
_{y}=\hat{I}_{y}\hat{\Pi}_{-},\qquad {(\hat{\Pi}_{+}\varphi )}^{\ast }=\hat{%
\Pi}_{-}\varphi ^{\ast }  \label{a2.23}
\end{equation}
It follows from Eqs. (\ref{a2.20}) -- (\ref{a2.23})
\begin{equation}
\varphi (x)=\hat{\Pi}_{+}\varphi _{{\rm E}}(x)+\hat{\Pi}_{-}\hat{I}%
_{0}\varphi _{{\rm E}}^{\ast }(x),  \label{a2.24}
\end{equation}
Thus, if $\varphi (x)$ is quantized according to $PA$-scheme, then $\varphi
_{{\rm E}}(x)$ is quantized according to $WL$-scheme.
\begin{equation}
{\lbrack \varphi (x),\varphi ^{\ast }(x^{\prime })]}_{-}=-iD(x-x^{\prime })
\label{a2.25}
\end{equation}
where $D(x-x^{\prime })$ is the Pauli-Jordan commutation function in the $%
(n+1)$-dimensional space-time.
\[
D(x)=\frac{1}{(2\pi )^{n}i}\int \frac{d{\bf k}}{\beta ({\bf k})}\left( e^{iE(%
{\bf k})x^{0}-i{\bf kx}}-e^{-iE({\bf k})x^{0}-i{\bf kx}}\right) =D^{+}\left(
x\right) +D^{-}\left( x\right)
\]
\[
D^{+}\left( x\right) =\frac{1}{(2\pi )^{n}i}\int \frac{d{\bf k}}{\beta ({\bf %
k})}e^{iE({\bf k})x^{0}-i{\bf kx}}
\]
\[
D^{-}(x)=\frac{i}{(2\pi )^{n}}\int \frac{d{\bf k}}{\beta ({\bf k})}e^{-iE(%
{\bf k})x^{0}-i{\bf kx}}=-D^{+}(-x)
\]
In this case $\varphi _{{\rm E}}(x)$ is quantized according to $WL$-scheme.
The commutation relation has the form
\begin{equation}
{\lbrack \varphi _{{\rm E}}(x),\varphi _{{\rm E}}^{\ast }(x^{\prime })]}%
_{-}=D_{1}(x-x^{\prime })=i[D^{+}(x-x^{\prime })-D^{-}(x-x^{\prime })]
\label{a2.26}
\end{equation}
For odd $n$ and $(x-x^{\prime })^{2}<0$ the condition $D(x-x^{\prime })=0$
is fulfilled. But the function $D_{1}(x-x^{\prime })$ has not this property.
It is a common practice to believe \cite{P41} that a violation of this
property leads to a causality violation. Producing a transformation from $PA$%
-scheme to $WL$-scheme, the change of designation (\ref{a2.17}) cannot
violate causality, as well other physical properties. Nevertheless it leads
to relation (\ref{a2.26}) which does not vanish at $(x-x^{\prime })^{2}<0.$
It means only that the this property is an attribute of the $PA$-scheme of
quantization (see also discussion of this question in ref.\cite{R72}).

Let $\varphi _{1}(x)$ and $\varphi _{2}(x)$ be two scalars. At the
translation
\begin{equation}
x^{i}\rightarrow \tilde{x}^{i}=x^{i}+a^{i},\qquad \tilde{x}=x+a,\qquad a^{i}=%
\hbox{const\qquad }  \label{a2.27}
\end{equation}
each of them transforms according to the law
\begin{equation}
\varphi (x)\rightarrow \tilde{\varphi}(\tilde{x})=\varphi (x)=\varphi (%
\tilde{x}+a)  \label{a2.28}
\end{equation}
At the proper Lorentz transformation
\begin{equation}
x^{i}\rightarrow \tilde{x}^{i}=\Lambda _{.k}^{i.}x^{k},\qquad \tilde{x}%
=\Lambda x,\qquad \Lambda _{.k}^{i.}g^{kl}\Lambda _{.l}^{j.}=g^{ij},\qquad
\Lambda _{.0}^{0.}>0  \label{a2.29}
\end{equation}
each of them transforms
\begin{equation}
\varphi (x)\rightarrow \tilde{\varphi}(\tilde{x})=\varphi (x)=\varphi
(\Lambda ^{-1}\tilde{x}),\qquad {(\Lambda \Lambda ^{-1})}_{.k}^{i.}=\delta
_{k}^{i}.  \label{a2.30}
\end{equation}
Then the quantity
\begin{equation}
f_{y}(x)=\varphi _{1}(x)+\hat{I}_{y}\varphi _{2}(x)=\varphi _{1}(x)+\varphi
_{2}(2y-x),  \label{a2.31}
\end{equation}
considered to be a function of argument $x$ and $y$ is a scalar
\begin{equation}
f_{y}(x)\rightarrow \tilde{f}_{\tilde{y}}(\tilde{x})=f_{y}(x).  \label{a2.32}
\end{equation}
At the same time the quantity $f_{y}(x)$ considered to be a function of only
$x$ is not a scalar. The quantity (\ref{a2.31}) considered to be a function
of only $x$ will be referred to as a scalar centaur with the contact point $%
y $.

Let the scalar $\varphi $ satisfy Eq. (\ref{a2.1}). Then $\hat{\Pi}%
_{+}\varphi (x)$ and $\hat{\Pi}_{+}\varphi ^{\ast }(x)$ are scalars
satisfying Eq.(\ref{a2.1}). The quantity $\varphi _{{\rm E}}(x)$ defined by
Eq.(\ref{a2.20}) is a scalar centaur with the contact point $y=0.$ This
scalar centaur $\varphi _{{\rm E}}(x)$ satisfies Eq.(\ref{a2.1}) also.

If a scalar field $\varphi (x)$ is quantized according to $PA$-scheme, then $%
\varphi _{{\rm E}}(x)$ is a scalar centaur which is quantized according to $%
WL$-scheme. If $\varphi _{{\rm E}}(x)$ is a scalar quantized according to $%
WL $-scheme, then $\varphi (x)$ defined by Eq.(\ref{a2.24}) is a scalar
centaur quantized according to $PA$-scheme.

Thus, $PA$-scheme and $WL$-scheme of quantization are equivalent in the
sense that each of them can be transformed into another by means of a change
of variables, but at such a transformation a scalar transforms into a scalar
centaur and vice versa.

Note that the transformation (\ref{a2.22}) associates with the
transformation (\ref{a1.16}), but not with the transformation (\ref{a1.15}),
because description of the field $\varphi $ does not contain a quantity
analogical to parameter $\tau $ in the action (\ref{a1.4}). Taking into
account the constraint (\ref{a1.19}), one is forced to use transformation
analogical to (\ref{a1.16}) and use different dynamic systems for
description of the field $\varphi $. In the absence of electromagnetic field
$\left( A_{i}=0\right) $ the action (\ref{a1.4}), (\ref{b1.5}) is invariant
with respect to transformation (\ref{a1.16}), and one can use this
transformation as well as the transformation (\ref{a2.12}) in the case of
free field $\varphi $. In the case of interaction, when $A_{i}\neq 0$, the
action (\ref{a1.4}), (\ref{b1.5}) is not invariant with respect to
transformation (\ref{a1.16}), and one cannot use this transformation. The
same relates to the non-linear field $\varphi (x)$, as we shall see in the
fourth section.

\section{Quantization Model}

{\it Definition} 3.1. The quantization model is a totality of the
quantization scheme and observable quantities represented as functions of
dynamic variables in this quantization scheme.

Different quantization models can exist at the same quantization scheme. Let
us introduce observable quantities by means of corresponding classic
quantities. The Lagrangian density has the form
\begin{equation}
L=\varphi _{i}^{\ast }\varphi ^{i}-m^{2}\varphi ^{\ast }\varphi ,\qquad
\varphi _{i}\equiv \partial _{i}\varphi ,\qquad \varphi ^{i}\equiv \partial
^{i}\varphi =g^{ik}\varphi _{k}  \label{a3.1}
\end{equation}
Let us define the energy $E$ and momentum $P_{\alpha }$ by relations
\[
P_{0}=E=\int T_{0.}^{.0}d{\bf x},\qquad P_{\alpha }=\int T_{\alpha .}^{.0}d%
{\bf x},
\]
\begin{equation}
T_{i.}^{.k}=\varphi _{i}^{\ast }\varphi ^{k}+\varphi ^{\ast k}\varphi
_{i}-\delta _{i}^{k}L  \label{a3.2}
\end{equation}
The number $N$ of SWLs and the electric charge $Q$ are defined by relations
\begin{equation}
N=-i\int (\dot{\varphi}^{\ast }\varphi -\varphi ^{\ast }\dot{\varphi})d{\bf x%
},\qquad Q=eN  \label{a3.3}
\end{equation}
In the expression for the energy-momentum $P_{i}$ and for the charge $Q$ the
normal ordering should be used, i.e. the creation operators should be placed
to the left of the annihilation ones. In the $WL$-scheme the operator $%
\varphi $ contains only annihilation operators, and $\varphi ^{\ast }$
contains only creation operators. Then the expressions (\ref{a3.1})-(\ref
{a3.3}) appear to be normally ordered automatically.

Let $\varphi (x)$ satisfy Eq.(\ref{a2.1}) and be quantized according to $PA$%
-scheme. Let $\varphi _{E}(x)$ be connected with $\varphi (x)$ by
transformation (\ref{a2.20}). Then a calculation leads to the following
expressions for $P_{i}$ and $Q$
\begin{equation}
P_{0}=E=\int E({\bf k})[c^{\ast }({\bf k})c({\bf k})+d^{\ast }({\bf k})d(%
{\bf k})]dk=\int E({\bf k})b_{{\rm E}}^{\ast }(K)b_{{\rm E}}(K)dK
\label{a3.4}
\end{equation}
\begin{equation}
P_{\alpha }=\int k_{\alpha }[c^{\ast }({\bf k})c({\bf k})-d^{\ast }({\bf k}%
)d({\bf k})]d{\bf k}=\int \varepsilon _{k}k_{\alpha }b_{{\rm E}}^{\ast
}(K)b_{{\rm E}}(K)dK  \label{a3.5}
\end{equation}
\begin{equation}
Q=\int e[c^{\ast }({\bf k})c({\bf k})-d^{\ast }({\bf k})d({\bf k})]d{\bf k}%
=\int e\varepsilon _{k}b_{{\rm E}}^{\ast }(K)b_{{\rm E}}(K)dK  \label{a3.6}
\end{equation}
Let $\varphi (x)$ satisfy Eq.(\ref{a2.1}) and be quantized according to $WL$%
-scheme. Then
\begin{equation}
P_{i}=\int \varepsilon _{k}k_{i}b^{\ast }(K)b(K)dK,\qquad k_{0}=\varepsilon
_{k}E({\bf k}),\qquad i=0,1,\ldots n  \label{a3.7}
\end{equation}
\begin{equation}
Q=\int e\varepsilon _{k}b^{\ast }(K)b(K)dK  \label{a3.8}
\end{equation}
Comparison of Eqs. (\ref{a3.4})-(\ref{a3.6}) with Eqs. (\ref{a3.7}, (\ref
{a3.8}) shows that at the quantization of $\varphi $ according to $PA$%
-scheme the expressions of integral quantities $P_{i},Q$ through operators $%
b_{E}^{\ast },b_{E}$ have the same form as the expressions of the same
quantities $P_{i}$, $Q$ through the operators $b^{\ast }$, $b$ at
quantization of $\varphi $ according to $WL$-scheme. Let us formulate this
circumstance as follows. The $PA$-model of the field $\varphi $ quantization
is equivalent to the $WL$-model of the field $\varphi $ with respect to
integral quantities $P_{i}$, $Q$.

But $PA$-model and $WL$-model are not equivalent with respect to local
quantities of the type of the current density $j^{i}$ or the energy momentum
tensor $T^{ik}$. For instance, at the quantization of $\varphi $ according
to $PA$-scheme one obtains
\[
PA:\qquad j^{0}=-ie:\dot{\varphi}^{\ast }(x)\varphi (x)-\varphi ^{\ast }(x)%
\dot{\varphi}(x):
\]
\[
=\frac{e}{(2\pi )}\int\limits_{\infty }^{\infty }{\frac{d{\bf k}d{\bf %
k^{\prime }}}{\sqrt{\beta ({\bf k})\beta ({\bf k}^{\prime })}}}\exp [-i({\bf %
k}-{\bf k^{\prime }}){\bf x}]
\]
\[
\times \{[E({\bf k})+E({\bf k^{\prime }})][b_{{\rm E}}^{\ast }(1,{\bf k}%
,x^{0})b_{{\rm E}}(1,{\bf k^{\prime }},x^{0})-b_{{\rm E}}^{\ast }(-1,{\bf %
k^{\prime }},x^{0})b_{{\rm E}}(-1,{\bf k},x^{0})]
\]
\begin{equation}
+[E({\bf k})-E({\bf k^{\prime }})][b_{{\rm E}}^{\ast }(1,{\bf k},x^{0})b_{%
{\rm E}}^{\ast }(-1,{\bf k^{\prime }},x^{0})-b_{{\rm E}}(-1,{\bf k},x^{0})b_{%
{\rm E}}(1,{\bf k^{\prime }},x^{0})]\}  \label{a3.9}
\end{equation}
Here colon '':'' denotes the normal ordering.

At the quantization of the field $\varphi $ according to $WL$-scheme one
obtains
\[
E:\qquad j^{0}=-ie[\dot{\varphi}^{\ast }(x)\varphi (x)-\varphi ^{\ast }(x)%
\dot{\varphi}(x)]
\]
\begin{eqnarray}
&=&\frac{e}{(2\pi )^{n}}\int\limits_{-\infty }^{\infty }\exp \{i[\varepsilon
_{k}E({\bf k})-\varepsilon _{k^{\prime }}E({\bf k}^{\prime })]x^{0}-i({\bf k}%
-{\bf k}^{\prime }){\bf x}\}  \nonumber \\
&&\times \frac{\varepsilon _{k}E({\bf k})+\varepsilon _{k^{\prime }}E({\bf k}%
^{\prime })}{\sqrt{\beta ({\bf k})\beta ({\bf k}^{\prime })}}b^{\ast
}(K)b(K^{\prime })dKdK^{\prime }  \label{a3.10}
\end{eqnarray}
Integration of Eqs. (\ref{a3.9}) and (\ref{a3.10}) over ${\bf x}$ leads to
the same result (\ref{a3.6}), (\ref{a3.8}).

Let us define canonical momentum $p_{i}=(-H,p_{\alpha })=(-H,-\pi _{\alpha
}) $ by means of relation
\begin{equation}
\partial _{k}\varphi (x)=i\left[ \varphi (x),p_{k}\right] _{-}  \label{a3.11}
\end{equation}
For $PA$-scheme one obtains from Eqs. (\ref{a3.11}), (\ref{a2.9}), (\ref
{a2.11}) and (\ref{a2.17})
\[
PA:\qquad H=-p_{0}=\int E({\bf k})[c^{\ast }({\bf k})c({\bf k})+d^{\ast }(%
{\bf k})d({\bf k})]d{\bf k}
\]
\begin{equation}
\pi _{\alpha }=-p_{\alpha }=\int k_{\alpha }[c^{\ast }({\bf k})c({\bf k}%
)-d^{\ast }({\bf k})d({\bf k})]d{\bf k,\qquad }\alpha =1,2,...n
\label{a3.12}
\end{equation}
At quantization according to $WL$-scheme it follows from Eqs. (\ref{a3.11}),
(\ref{a2.13}), (\ref{a2.15})
\begin{equation}
WL:\qquad \pi _{i}=-p_{i}=\int k_{i}b^{\ast }(K)b(K)dK,\qquad i=0,1,...n
\label{a3.13}
\end{equation}
Comparing Eq. (\ref{a3.12}) with Eqs. (\ref{a3.4}), (\ref{a3.5}), one can
see that Eq.(\ref{a1.18}) is fulfilled. It is this condition that is a
criterion of $PA$-approach.

One can see from (\ref{a3.13}) and (\ref{a3.8}) that the condition (\ref
{a1.18}) is fulfilled for contribution with $\varepsilon _{k}=1,$ and it is
not fulfilled for contribution with $\varepsilon _{k}=-1.$

In the classical case a transformation from $WL$-approach to $PA$-approach
can be realized by means of transformation (\ref{a1.15}), where the
canonical momentum reflection is produced without the coordinate reflection.
In the quantum theory for a transition from $WL$-approach to $PA$-approach
one has to use the transformation (\ref{a1.16}), where the canonical
momentum reflection is accompanied with the coordinate reflection. It leads
to transformation of the scalar into the centaur.

Neither consideration of the quantization scheme, nor that of the
quantization model of the linear scalar field answers the question which of
the two approaches is valid. But such a question should be put, because both
approaches cannot be valid simultaneously. If $\varphi (x)$ is a scalar and
the quantization according to $WL$-scheme is valid, then the quantization
according to $PA$-scheme cannot be valid, because in this case $\varphi (x)$
is a centaur. Vice versa, if the quantization of the scalar field $\varphi
(x)$ according to $PA$-scheme is valid, then its quantization according to $%
WL$-scheme cannot be valid, because in this case $\varphi (x)$ is a centaur.
Essentially, the problem is reduced to the question how to distinguish
between a scalar and a centaur, as far as the centaur is a "spoiled" scalar.

\section{Quantization of Nonlinear Scalar Field}

Let us consider a quantization of the charged scalar field $\varphi (x)$
described by the Lagrangian density
\begin{equation}
L=:\varphi _{i}^{\ast }\varphi ^{i}-m^{2}\varphi ^{\ast }\varphi +{\frac{%
\lambda }{2}}\varphi ^{\ast }\varphi ^{\ast }\varphi \varphi :  \label{a4.1}
\end{equation}
\[
\varphi =\varphi (x),\qquad \varphi _{i}\equiv \partial _{i}\varphi ,\qquad
\varphi ^{i}\equiv \partial ^{i}\varphi ,\qquad x=(t,x).
\]
Here $\lambda $ is a self-action constant.

Let us introduce variables $b(K,t),b^{\ast }(K,t)$ by means of relations (%
\ref{a2.2}) - (\ref{a2.6}). Then dynamic equation for the field $\varphi $.
\begin{equation}
(\partial _{i}\partial ^{i}+m^{2})\varphi =:\lambda \varphi ^{\ast }\varphi
\varphi :  \label{a4.2}
\end{equation}
transforms into the equation
\begin{equation}
\dot{b}(K,t)=-i\varepsilon _{k}E({\bf k})b(K,t)+:i\lambda {\frac{\varepsilon
_{k}g({\bf k},t)}{\sqrt{\beta ({\bf k})}}}:  \label{a4.3}
\end{equation}
where
\begin{equation}
g({\bf k},t)=\frac{1}{(2\pi )^{n}}\int \int \int {\frac{\delta ({\bf k}+{\bf %
p}-{\bf p^{\prime }}-{\bf k^{\prime }})}{\sqrt{\beta ({\bf p})\beta ({\bf %
k^{\prime }})\beta ({\bf p^{\prime }})}}}b^{\ast }(P,t)b(P^{\prime
},t)b(K^{\prime },t)dPdP^{\prime }dK^{\prime }.  \label{a4.4}
\end{equation}
The colon denotes the normal ordering. An expression between two colons is
considered as normally ordered. The field is scalar with respect to a
transformation of the Poincar\'{e} group, i.e. it transforms according to
Eqs. (\ref{a2.27}) - (\ref{a2.30})

Let the field $\varphi $ be quantized according to $WL$-scheme. In this case
the colon can be omitted in Eq.(\ref{a4.3}). The unique vacuum state vector $%
\Phi _{0}=|0\rangle ,\Phi _{0}^{\ast }=\langle 0|$ is supposed to exist. It
is defined by relations
\begin{equation}
b(K)|0\rangle =0,\qquad \langle 0|b^{\ast }(K)=0,\qquad b(K)=b(K,0).
\label{a4.5}
\end{equation}
It follows from Eqs.(\ref{a4.3}), (\ref{a4.5})
\begin{equation}
\dot{b}(K)|0\rangle =0,\qquad \langle 0|\dot{b}^{\ast }(K)=0  \label{a4.6}
\end{equation}
and, hence
\begin{equation}
b(K)|0\rangle =0,\qquad \varphi (x)|0\rangle =0,\qquad \forall x\in {\Bbb R}%
^{n+1},  \label{a4.7}
\end{equation}
\begin{equation}
\partial _{k}\varphi (x)|0\rangle =0,\qquad \forall x\in {\Bbb R}^{n+1},
\label{a4.8}
\end{equation}

From Eqs.(\ref{a4.7}), (\ref{a4.8}) and definition (\ref{a3.11}) of the
canonical momentum operator it follows
\begin{equation}
\partial _{k}\varphi (x)|0\rangle =i\varphi (x)p_{k}|0\rangle =0,\qquad
\forall x\in {\Bbb R}^{n+1},  \label{a4.9}
\end{equation}
As far as only one vacuum state defined by Eq.(\ref{a4.5}), or by (\ref{a4.8}%
) exists, it follows from Eq.(\ref{a4.9}) that $p_{k}|0\rangle $
distinguishes from $|0\rangle $ by a factor only. It means the vacuum $%
|0\rangle $ is an eigenvector of the operator $p_{k}$
\begin{equation}
p_{k}|0\rangle =p_{k}^{\prime }|0\rangle  \label{a4.10}
\end{equation}
where $p_{k}^{\prime }$, $(k=0,1,...n)$ are $c$-numbers.

The vacuum state $|0\rangle $ is supposed to be invariant with respect to
Poincar\`{e} group transformations (\ref{a2.27})-(\ref{a2.30}). This
supposition is compatible with Eq.(\ref{a4.7}), (\ref{a4.8}) and dynamic
equations. In particular, the vacuum state $|0\rangle $ is stationary, as it
follows from Eq.(\ref{a4.10}).

Now let the field $\varphi (x)$ be quantized according to $PA$-scheme. Then
according to Eqs. (\ref{a2.12}), (\ref{a2.17})

\begin{equation}
b(\varepsilon _{k},{\bf k},t)=b_{{\rm E}}^{-\varepsilon _{k}}(\varepsilon
_{k},{\bf k},\varepsilon _{k}t)  \label{a4.11}
\end{equation}
where the following designation is used

\begin{equation}
b_{{\rm E}}^{\varepsilon }(K,t)=\left\{
\begin{array}{lll}
b_{{\rm E}}^{\ast }(K,t), & \hbox{  if } & \varepsilon =1 \\
b_{{\rm E}}(K,t), & \hbox{  if } & \varepsilon =-1
\end{array}
\right.  \label{a4.12}
\end{equation}
In terms of variables $b_{{\rm E}}$ Eq.(\ref{a4.4}) takes the form
\[
\dot{b}_{{\rm E}}(K,t)=-i\varepsilon _{k}E({\bf k})b_{{\rm E}}(K,t)+\frac{%
i\lambda \varepsilon _{k}}{(2\pi )^{n}}\int \int \int {\frac{\delta ({\bf k}+%
{\bf p}-{\bf q}-{\bf r})}{\sqrt{\beta ({\bf k})\beta ({\bf p})\beta ({\bf q}%
)\beta ({\bf r})}}}
\]
\begin{equation}
\times :b_{{\rm E}}^{-\varepsilon _{k}\varepsilon _{p}}(\varepsilon _{p},%
{\bf p},\varepsilon _{k}\varepsilon _{p}t)b_{{\rm E}}^{\varepsilon
_{k}\varepsilon _{q}}(\varepsilon _{q},{\bf q},\varepsilon _{k}\varepsilon
_{q}t)b_{{\rm E}}^{\varepsilon _{k}\varepsilon _{r}}(\varepsilon _{r},{\bf r}%
,\varepsilon _{k}\varepsilon _{r}t):dPdQdR  \label{a4.13}
\end{equation}

Definition of the vacuum vector $|0\rangle $ has the form of the second
relation of Eq.(\ref{a2.18}). At $t=0$ it follows from Eq. (\ref{a4.13}) and
\begin{equation}
\dot{b}_{{\rm E}}(K)|0\rangle _{{\rm PA}}=\frac{i\lambda \varepsilon _{k}}{%
(2\pi )^{n}}\int \int \int \frac{b_{{\rm E}}^{\ast }(-\varepsilon _{k},{\bf p%
})b_{{\rm E}}^{\ast }(\varepsilon _{k},{\bf q})b_{{\rm E}}^{\ast
}(\varepsilon _{k},{\bf k}+{\bf p}-{\bf q})}{\sqrt{\beta ({\bf k})\beta (%
{\bf p})\beta ({\bf q})\beta ({\bf k}+{\bf p}-{\bf q})}}|0\rangle _{{\rm PA}%
}d{\bf p}d{\bf q}  \label{a4.14}
\end{equation}
Thus, generally,
\begin{equation}
\dot{b}_{{\rm E}}(K)|0\rangle _{{\rm PA}}\neq 0,  \label{a4.15}
\end{equation}

It means the vacuum vector $|0\rangle $ cannot be an eigenvector of the
Hamiltonian $H=-p_{0}$, i.e. $|0\rangle _{{\rm PA}}$ is not a stationary
vector. The fact that the vacuum $|0\rangle _{{\rm PA}}$ is nonstationary
excites a disappointment and dissatisfaction, because it means a translation
non-invariance of vacuum $|0\rangle _{PA}$ (see discussion of this question
in sec. 6 of ref.[7]) and that of the $PA$-scheme of quantization. In
application to Eq.(\ref{a2.7}) the $PA$- and $WL$-scheme are equivalent, but
in application to Eqs. (\ref{a4.3}), (\ref{a4.4}) they are not equivalent,
because the form of Eq.(\ref{a4.3}) is not invariant with respect to
transformation (\ref{a4.11}). If one needs to select between the two
schemes, then it is reasonable to select the scheme, where the vacuum state
is well defined.

Thus, the $WL$-scheme is selected. Further we shall deal with the scalar
field quantized according to $WL$-scheme.

The commutation relation between dynamic variables $b(K,t)$ and $b^{\ast
}(K^{\prime },t)$ are necessary for a calculation of matrix elements of an
operator ${\cal R}$ between two states $\Phi $ and $\Phi ^{\prime }$. Any
state $\Phi $ can be represented as follows
\begin{equation}
\Phi _{l}=\sum\limits_{l=0}\Phi _{l},\qquad \Phi _{l}^{\ast
}=\sum\limits_{l=0}\Phi _{l}^{\ast },  \label{a4.16}
\end{equation}
where $\Phi _{l}$ is an $l$-WL state defined by the relations
\[
\Phi _{l}=\frac{1}{\sqrt{l!}}\int f_{l}({\cal K}^{l})B_{l}^{\ast }({\cal K}%
^{l})|0\rangle d{\cal K}^{l}\equiv \int f_{l}({\cal K}^{l})|{\cal K}%
^{l}\rangle d{\cal K}^{l}
\]
\begin{equation}
\Phi _{l}^{\ast }=\frac{1}{\sqrt{l!}}\int \langle 0|f_{l}^{\ast }({\cal K}%
^{l})B_{l}({\cal K}^{l})d{\cal K}^{l}\equiv \int \langle {\cal K}%
^{l}|f_{l}^{\ast }({\cal K}^{l})d{\cal K}^{l}  \label{a4.17}
\end{equation}
where
\[
{\cal K}^{l}\equiv \{K_{1},K_{2},\ldots K_{l}\},\qquad d{\cal K}^{l}\equiv
dK_{1}dK_{2}\ldots dK_{l}
\]
\begin{equation}
B_{l}^{\ast }({\cal K}^{l})\equiv b^{\ast }(K_{1})b^{\ast }(K_{2})\ldots
b^{\ast }(K_{l})  \label{a4.18}
\end{equation}
\[
B_{l}({\cal K}^{l})\equiv b(K_{l})b(K_{l-1})\ldots b_{1}(K)
\]
and $f_{l}({\cal K}^{l})$ is a complex function of arguments $%
K_{1},K_{2},\ldots K_{l}$. The wave function $f_{l}({\cal K}^{l})$ is
symmetric, if it does not change at transposition of any two arguments. The
wave function is antisymmetric, if it changes sign at transposition of any
two arguments.

For calculation of $(\Phi ,{\cal R}\Phi )$ it is sufficient to know
commutation relations
\begin{equation}
\lbrack b(K,t),b^{\ast }(K^{\prime },t^{\prime })]_{-}=D(t,t^{\prime
};K,K^{\prime })  \label{a4.19}
\end{equation}
\begin{equation}
b(K,t)b^{\ast }(K^{\prime },t^{\prime })=F(t,t^{\prime };K,K^{\prime })
\label{a4.20}
\end{equation}
where $D(t,t^{\prime };K,K^{\prime })$ and $F(t,t^{\prime };K,K^{\prime })$
are operators depending on parameters $K$, $K^{\prime }$, $t$, $t^{\prime }$%
. They are supposed to be functions of dynamic variables $b(K),b^{\ast }(K)$
disposed in the normal order, when in each term any creation operator $%
b^{\ast }(K)$ is placed to the left of all annihilation operators $b(K)$. $%
D(t,t^{\prime };K,K^{\prime })$ and $F(t,t^{\prime };K,K^{\prime })$ are
connected by the evident relation
\begin{equation}
F(t,t^{\prime };K,K^{\prime })=D(t,t^{\prime };K,K^{\prime })+b^{\ast
}(K^{\prime },t^{\prime })b(K,t)  \label{a4.21}
\end{equation}
Determination of the operators $D$ and $F$ is equivalent to a solution of
the equation (\ref{a4.3}). For calculating matrix elements the commutation
relations $\left[ b(K,t),b(K,t^{\prime })\right] _{-}$, $\left[ b^{\ast
}(K,t),b^{\ast }(K^{\prime },t^{\prime })\right] _{-}$ are not necessary.
These commutation relations are important at the WL identity consideration.

For determination of operators $D$ and $F$ the following propositions are
supposed to be fulfilled.

\noindent I. Relations (\ref{a4.19}), (\ref{a4.20}) have all symmetries that
the field $\varphi $ has.

\noindent II. Relations (\ref{a4.19}), (\ref{a4.20}) are compatible with the
dynamic equation (\ref{a4.3}).

\noindent III. Relations (\ref{a4.19}), (\ref{a4.20}) are invariant with
respect to Poincare group.

\noindent IV. For any $l$-WL state (\ref{a4.17}) $l=1,2,\ldots $ the scalar
product $(\Phi ,\Phi )>0$, if the symmetrical part of the wave function $%
f_{l}({\cal K}^{l})$ does not vanish, and $(\Phi ,\Phi )=0,$ if the wave
function $f_{l}({\cal K}^{l})$ is antisymmetric.

\noindent V. The operator (\ref{a3.8}) has only $e$-fold eigenvalues. At the
$l$-WL state the eigenvalues of $Q/e$ are equal to $-l, -l+2,\ldots l$.

\noindent VI. At $t=t^{\prime }$ in the limit $\lambda \rightarrow 0$ the
commutation relations (\ref{a4.19}) turn to Eq.(\ref{a2.15}).

\noindent It follows from Eq. (\ref{a4.20})

\begin{equation}
F(t,t^{\prime };K,K^{\prime })=F^{\ast }(t^{\prime },t;K^{\prime },K)
\label{a4.22}
\end{equation}
As far as lhs of Eq.(\ref{a4.20}) is invariant with respect to transformation

\[
b(K,t)\rightarrow \tilde{b}(K,t)=b(K,t)e^{i\alpha },\qquad \alpha =%
\hbox{const}
\]
\begin{equation}
b^{\ast }(K,t)\rightarrow \tilde{b}^{\ast }(K,t)=b^{\ast }(K,t)e^{-i\alpha },
\label{a4.23}
\end{equation}
the operator $F$ has the form
\[
F(t,t^{\prime };K,K^{\prime })=F_{0}(t,t^{\prime };K,K^{\prime })+
\]
\begin{equation}
\sum\limits_{l=1}^{\infty }\int \int F_{l}(t,t^{\prime };K,K^{\prime };{\cal %
P}^{l},{\cal P}^{\prime l})B_{l}^{\ast }({\cal P}^{l})B_{l}({\cal P}^{\prime
l})d{\cal P}d{\cal P}^{\prime l}  \label{a4.24}
\end{equation}
where $F_{l}(t,t^{\prime };K,K^{\prime };{\cal P}^{l},{\cal P}^{\prime
l})\quad l=0,1,2,\ldots $ are some $c$-numerical functions of their
arguments. The form of functions $F_{l}$ is determined by the conditions
II-VI. The operator $D$ expansion has a like form.

Differentiating Eq.(\ref{a4.20}) with respect to $t$ and using Eq. (\ref
{a4.3}), one obtains an integro-differential equation for $F(t,t^{\prime
};K,K^{\prime })$

\[
\frac{\partial F}{\partial t}(t,t^{\prime };K,K^{\prime })=-i\varepsilon
_kE(k)F(t,t^{\prime };K,K^{\prime })
\]
\begin{equation}
+\frac{i\lambda \varepsilon _k}{(2\pi )^n}\int \int \int {\frac{\delta ({\bf %
k}+{\bf p}-{\bf k^{\prime \prime }}-{\bf p^{\prime \prime }})}{\sqrt{\beta (%
{\bf k})\beta ({\bf p})\beta ({\bf k^{\prime \prime }})\beta ({\bf p^{\prime
\prime }})}}}b^{*}(P,t)b(P^{\prime \prime },t)F(t,t^{\prime };K^{\prime
\prime },K^{\prime })dK^{\prime \prime }dP^{\prime \prime }dP  \label{a4.25}
\end{equation}
The expression for $\partial F/\partial t^{\prime }$ is obtained from Eq.(%
\ref{a4.25}) by means of Hermitian conjugation and the substitution $%
t\leftrightarrow t^{\prime }$, $K\leftrightarrow K^{\prime }$.

Let us take the matrix element $\langle 0|\ldots |0\rangle $ from Eq.(\ref
{a4.25}) and a like equation for $\partial F/\partial t^{\prime }$. One
obtains
\[
\frac{\partial F_{0}}{\partial t}(t,t^{\prime };K,K^{\prime })=-i\varepsilon
_{k}E(k)F_{0}(t,t^{\prime };K,K^{\prime })
\]
\begin{equation}
\frac{\partial F_{0}}{\partial t^{\prime }}(t,t^{\prime };K,K^{\prime
})=i\varepsilon _{k^{\prime }}E(k)F_{0}(t,t^{\prime };K,K^{\prime })
\label{a4.26}
\end{equation}
Taking into account conditions III$-V$, the solution of these equations are
unique and has the form [2]
\begin{equation}
D_{0}(t,t^{\prime };K,K^{\prime })=F_{0}(t,t^{\prime };K,K^{\prime })=\delta
(K-K^{\prime })e^{-i\varepsilon _{k}E(k)(t-t^{\prime })}  \label{a4.27}
\end{equation}
It agrees with the commutation relation (\ref{a2.15}) for the free field,
quantized according to $WL$-scheme.

Let us take matrix element $\langle P|\ldots |P^{\prime }\rangle $ from Eq. (%
\ref{a4.25}). Then using Eq.(\ref{a4.27}), one obtains
\[
\frac{\partial F_{1}}{\partial t}(t,t^{\prime };K,K^{\prime };P,P^{\prime
})=-i\varepsilon _{k}E(k)F_{1}(t,t^{\prime };K,K^{\prime };P,P^{\prime })
\]
\[
+\frac{i\lambda \varepsilon _{k}}{(2\pi )^{n}}\int \int \int {\frac{\delta (%
{\bf k}+{\bf p}-{\bf k^{\prime \prime }}-{\bf p^{\prime \prime }})}{\sqrt{%
\beta ({\bf k})\beta ({\bf p})\beta ({\bf k^{\prime \prime }})\beta ({\bf %
p^{\prime \prime }})}}}e^{i[\varepsilon _{p}E({\bf p})-\varepsilon
_{p^{\prime \prime }}E({\bf p^{\prime \prime }})]t}
\]
\begin{equation}
\times F_{1}(t,t^{\prime };K^{\prime \prime },K^{\prime };P^{\prime \prime
},P^{\prime })dK^{\prime \prime }dP^{\prime \prime }  \label{a4.28}
\end{equation}
Calculating matrix elements $\langle {\cal P}^{l}|\ldots |{\cal P}^{\prime
}{}^{l}\rangle $ $(l=2,3,\ldots )$ from Eq.(\ref{a4.25}), one obtains a
chain of integro-differential equations for determination of the functions $%
F_{2},F_{3},\ldots $ The equation of the $l$th order is linear with respect
to functions $F_{l}$. Coefficients of this equation contain the functions $%
F_{l-1},F_{l-2}$,... which can be determined from the preceding equations of
the chain.

It is convenient to use the following designations
\begin{eqnarray}
-w &=&\{P,K\}\equiv \{{\bf s},-u\},\qquad u=\{\varepsilon _{p},\varepsilon
_{k},-{\bf q}/2\},\qquad {\bf q}={\bf k}-{\bf p}  \nonumber \\
w &=&\{K,P\}\equiv \{{\bf s},u\},\qquad u=\{\varepsilon _{k},\varepsilon
_{p},{\bf q}/2\},\qquad {\bf s}={\bf k}+{\bf p}  \label{a4.29}
\end{eqnarray}
\begin{eqnarray}
-w^{\prime } &=&\{P^{\prime },K^{\prime }\}\equiv \{{\bf s^{\prime }}%
,-u^{\prime }\},\qquad u^{\prime }=\{\varepsilon _{p^{\prime }},\varepsilon
_{k^{\prime }},-{\bf q^{\prime }}/2\},\qquad {\bf q^{\prime }}={\bf %
k^{\prime }}-{\bf p^{\prime }}  \nonumber \\
w^{\prime } &=&\{K^{\prime },P^{\prime }\}\equiv \{{\bf s^{\prime }}%
,u^{\prime }\},\qquad u^{\prime }=\{\varepsilon _{k^{\prime }},\varepsilon
_{p^{\prime }},{\bf q^{\prime }}/2\},\qquad {\bf s^{\prime }}={\bf k^{\prime
}}+{\bf p^{\prime }}  \label{a4.30}
\end{eqnarray}
\begin{eqnarray}
\int (.)dw &\equiv &\int \int (.)dKdP\equiv \int \int (.)d{\bf s}du,
\nonumber \\
\int (.)du &\equiv &2^{-n}\sum_{\varepsilon _{k},\varepsilon _{p}=\mp
1}\int\limits_{-\infty }^{\infty }(.)d{\bf q}  \label{a4.31}
\end{eqnarray}
\[
\delta (w-w^{\prime })\equiv \delta (K-K^{\prime })\delta (P-P^{\prime
})\equiv \delta ({\bf s}-{\bf s^{\prime }})\delta (u-u^{\prime })
\]
\begin{equation}
\delta (w+w^{\prime })\equiv \delta (K-P^{\prime })\delta (P-K^{\prime
})\equiv \delta ({\bf s}-{\bf s^{\prime }})\delta (u+u^{\prime })
\label{a4.32}
\end{equation}
Supposing
\[
F_{1}(t,t^{\prime };K,K^{\prime },P,P^{\prime })=e^{-i\varepsilon _{k}E({\bf %
k})t+i\varepsilon _{k^{\prime }}E({\bf k^{\prime }})t^{\prime
}}D_{1}^{\prime }(t,t^{\prime };w,w^{\prime })
\]
\begin{equation}
-e^{-i\varepsilon _{p}E({\bf p})(t-t^{\prime })}\delta (w-w^{\prime })
\label{a4.33}
\end{equation}
one obtains from Eq.(\ref{a4.28})
\[
\frac{\partial D_{1}^{\prime }}{\partial t}(t,t^{\prime };w,w^{\prime
})=i\lambda \int {\cal K}(w,w^{\prime \prime })e^{i[\omega (w)-\omega
(w^{\prime \prime }))]t}D_{1}^{\prime }(t,t^{\prime };w^{\prime \prime
},w^{\prime })dw^{\prime \prime }
\]
\begin{equation}
\frac{\partial D_{1}^{\prime }}{\partial t^{\prime }}(t,t^{\prime
};w,w^{\prime })=-i\lambda \int {\cal K}(w^{\prime },w^{\prime \prime
})e^{i[\omega (w^{\prime })-\omega (w^{\prime \prime })]t}D_{1}^{\prime
}(t,t^{\prime };w,w^{\prime \prime })dw^{\prime \prime }  \label{a4.34}
\end{equation}
where the following designations are used
\begin{equation}
{\cal K}(w,w^{\prime })=(2\pi )^{-n}\eta _{1}(w)\eta _{2}(w^{\prime })\delta
({\bf s}-{\bf s^{\prime }}),\qquad \eta _{1}(w)\equiv \zeta _{1}(u,{\bf s})={%
\frac{\varepsilon _{k}}{\sqrt{\beta ({\bf k})\beta ({\bf p})}}},
\label{a4.35}
\end{equation}
\begin{equation}
\eta _{2}(w)\equiv \zeta _{2}(u,{\bf s})={\frac{1}{\sqrt{\beta ({\bf k}%
)\beta ({\bf p})}}},\qquad \omega (w)=\omega _{1}(u,{\bf s})\equiv
\varepsilon _{k}E({\bf k})+\varepsilon _{p}E({\bf p})  \label{a4.36}
\end{equation}
The solution of Eq.(\ref{a4.34}) can be obtained in the form
\begin{equation}
D_{1}^{\prime }(t,t^{\prime };w,w^{\prime })=\int \int \kappa _{\overline{w}%
}^{(+)}(w)e^{i[\omega (\overline{w})-\omega (w)]t}B_{\overline{w}\overline{w}%
^{\prime }}\kappa _{\overline{w}}^{(-)}(w^{\prime })e^{-i[\omega (\overline{w%
}^{\prime })-\omega (w^{\prime })]t^{\prime }}d\overline{w}d\overline{w}%
^{\prime }  \label{a4.37}
\end{equation}
Here $\overline{w}$ and $\overline{w}^{\prime }$ label the eigenfunctions $%
\kappa _{\overline{w}}^{(\pm )}(w)$ of the equation
\begin{equation}
\lbrack \omega (\overline{w})-\omega (w)]\kappa _{\overline{w}}^{(\pm
)}(w)+\lambda \int {\cal K}(w,w^{\prime })\kappa _{\overline{w}}^{(\pm
)}(w^{\prime })dw^{\prime }=0  \label{a4.38}
\end{equation}
and $B_{\overline{w}\overline{w}^{\prime }}$is some function of arguments $%
\overline{w},\overline{w}^{\prime }$. Integral over $\overline{w}$ includes
summation over discrete spectrum and integration over continuous one.

For determination of eigenfunctions $\kappa ^{(\pm )}$ the following
integrals are important
\begin{equation}
I^{(\pm )}(\omega ,{\bf s})=\frac 1{(2\pi )^n}\int \frac{\eta _1(w)\eta
_2(w)\delta ({\bf s}-{\bf k}-{\bf p})}{\omega -\omega (w)\pm i0}dw
\label{a4.39}
\end{equation}
Here the symbol $\pm 0$ determines the round way of the integrand poles. It
means $\pm i\varepsilon $, $\varepsilon \rightarrow +0.$

As it was shown in ref.[2], the integrals $I^{(\pm )}(\omega ,{\bf s})$
depend on the only argument $z=\gamma ^{2}=\omega ^{2}-s^{2}$. They are
analytical functions of $z$ on the complex plane with a cut along the real
axis $[4m^{2},+\infty )$. $I^{(\pm )}$ are values of the integral $I(\gamma
) $ at the different cut edges.

Let us set
\begin{equation}
\Delta (z)=1+\lambda I(\omega ,{\bf s}),\qquad z=\omega ^{2}-s^{2}=\gamma
^{2}  \label{a4.40}
\end{equation}
The calculation [2] gives the following expression of $\Delta (z)$ at
different values of the dimension $n$ of the configurational space.
\begin{equation}
n=1,\qquad \Delta _{1}(z)=1-\frac{\lambda }{\pi \gamma \sqrt{4m^{2}-\gamma
^{2}}}\arctan \frac{\gamma }{\sqrt{4m^{2}-\gamma ^{2}}},\qquad \gamma =\sqrt{%
z}  \label{a4.41}
\end{equation}
\begin{equation}
n=2,\qquad \Delta _{2}(z)=1-{\frac{\lambda }{8\pi \gamma }}\log \frac{%
2m+\gamma }{2m-\gamma },\qquad \gamma =\sqrt{z}  \label{a4.42}
\end{equation}
For $n=3$ the integral (\ref{a4.39}) diverges. It can be represented in the
form
\begin{equation}
\Delta _{3}(z)=1-{\frac{\lambda }{8\pi ^{2}}}\left\{ \frac{\sqrt{%
4m^{2}-\gamma ^{2}}}{\gamma }\arctan \frac{\gamma }{\sqrt{4m^{2}-\gamma ^{2}}%
}+\lim_{\mu \rightarrow +\infty }\log \frac{\mu +\sqrt{4m^{2}+\mu ^{2}}}{2m}%
\right\}  \label{a4.43}
\end{equation}
The values of frequencies $\omega =\pm \sqrt{4m^{2}+{\bf s}^{2}}\geq 0$ at
the cut edges determine the continuous spectrum of Eq.(\ref{a4.38}). The
roots $z=M^{2}$ of the function $\Delta (z)$ determine the discrete spectrum
$\omega =\pm \sqrt{4m^{2}+{\bf s}^{2}}$. The analysis shows \cite{R72} $%
\Delta (z)$ has no roots, or one root depending on the value of $\lambda $.
The root lies in the interval $0\leq M^{2}<4m^{2}$.

Parameter $\overline{w}$ labelling the functions $\kappa _{\overline{w}%
}^{(\pm )}(w)$ can be represented as follows
\begin{equation}
\overline{w}=\{(\overline{K},\overline{P}),\overline{S}\},\qquad \overline{S}%
=\{\varepsilon _{\overline{s}},{\bf \overline{s}\}},\qquad \varepsilon _{%
\overline{s}}=\pm 1  \label{a4.44}
\end{equation}
\begin{equation}
\omega (\overline{w})=\{\omega (\overline{K},\overline{P}),\omega (\overline{%
S})\},\qquad \omega (\overline{K},\overline{P})=\varepsilon _{\overline{k}}E(%
\overline{{\bf k}})+\varepsilon _{\overline{p}}E(\overline{{\bf p}})\quad %
\hbox{for }\overline{w}=(\overline{K},\overline{P})  \label{a4.45}
\end{equation}
\begin{equation}
\omega (\overline{w})=\omega (\overline{S})=\varepsilon _{\overline{s}}E_{M}(%
{\bf \overline{s}})\quad \hbox{for\quad }\overline{w}=\overline{S}%
=\{\varepsilon _{\overline{s}},{\bf \overline{s}\}}  \label{a4.46}
\end{equation}
where
\begin{equation}
\beta _{M}({\bf s})=2E_{M}({\bf s})=2\left| \sqrt{M^{2}+{\bf s}^{2}}\right|
,\qquad \varepsilon _{s}=\pm 1  \label{a4.47}
\end{equation}
$M^{2}$ is a root of the function $\Delta (z)$, and $\varepsilon _{s}=\pm 1$
labels the states of the discrete spectrum.

For $\overline{w}=\{\overline{K},\overline{P}\}$ the eigenfunctions of Eq.(%
\ref{a4.38}) have the form
\begin{equation}
\kappa _{\overline{w}}^{(\pm )}(w)=\kappa _{\overline{K}\overline{P}}^{(\pm
)}(K,P)=\delta (w-\overline{w})-\frac{\lambda }{(2\pi )^{n}}\frac{\eta
_{1}(w)\eta _{2}(\overline{w})\delta ({\bf s}-{\bf \overline{s}})}{[\omega (%
\overline{w})-\omega (w)\pm i0]\Delta _{\pm }(\overline{w})}  \label{a4.48}
\end{equation}
where
\begin{equation}
\Delta _{\pm }(\overline{w})=1+\lambda I^{(\pm )}(\omega (\overline{w}),{\bf
\overline{s}}),\qquad {\bf \overline{s}}={\bf \overline{k}}+{\bf \overline{p}%
}  \label{a4.49}
\end{equation}
For discrete spectrum $\overline{w}=\overline{S}$, one has
\begin{equation}
\kappa _{\overline{w}}^{(\pm )}(w)=\kappa _{\overline{S}}^{(\pm )}(K,P)=%
\frac{1}{\sqrt{(2\pi )^{n}B}}\frac{\varepsilon _{s}\eta _{2}(w)\delta ({\bf s%
}-{\bf \overline{s}})}{\sqrt{\beta _{M}({\bf s})}[\omega (\overline{S}%
)-\omega (w)]}  \label{a4.50}
\end{equation}
\begin{equation}
B=\left| \frac{1}{\lambda }\frac{\partial \Delta (z)}{\partial z}\right|
_{z=M^{2}}  \label{a4.51}
\end{equation}
\noindent Equation (\ref{a4.38}) is not self-adjoint. The adjoint equation
has the form
\begin{equation}
\lbrack \omega (\overline{w})-\omega (w)]\overline{\kappa }_{\overline{w}%
}^{(\pm )}(w)+\lambda \int {\cal K}(w^{\prime },w)\overline{\kappa }_{%
\overline{w}}^{(\pm )}(w^{\prime })dw^{\prime }=0  \label{a4.52}
\end{equation}
Here $\overline{\kappa }_{\overline{w}}^{(\pm )}(w)$ are the functions
adjoint to $\kappa _{\overline{w}}^{(\pm )}(w)$. $\overline{\kappa }_{%
\overline{w}}^{(\pm )}(w)$ are obtained from $\kappa _{\overline{w}}^{(\pm
)}(w)$ by replacing $\eta _{1}\leftrightarrow \eta _{2}$, i.e.
\[
\overline{\kappa }_{\overline{w}}^{(\pm )}(w)=\varepsilon _{\overline{k}%
}\varepsilon _{k}\kappa _{\overline{w}}^{(\pm )}(w)\quad \hbox{for\quad }%
\overline{w}=(\overline{K},\overline{P})
\]
\begin{equation}
\overline{\kappa }_{\overline{S}}^{(\pm )}(w)=\varepsilon _{\overline{S}%
}\varepsilon _{k}\kappa _{\overline{w}}^{(\pm )}(w)\quad \hbox{for\quad }%
\overline{w}=\overline{S}  \label{a4.53}
\end{equation}
Each of the sets $\kappa _{\overline{w}}^{(+)}$ and $\kappa _{\overline{w}%
}^{(-)}$ is complete. They satisfy the relations
\begin{equation}
\int \overline{\kappa }_{\overline{w}}^{(\pm )}(w)\kappa _{\overline{w}%
^{\prime }}^{(\mp )}(w)dw=\delta (\overline{w}-\overline{w}^{\prime })
\label{a4.54}
\end{equation}
\begin{equation}
\int \overline{\kappa }_{\overline{w}}^{(\pm )}(w)\kappa _{\overline{w}%
}^{(\mp )}(w^{\prime })d\overline{w}=\delta (w-w^{\prime })  \label{a4.55}
\end{equation}
where
\begin{equation}
\delta (\overline{w}-\overline{w}^{\prime })=\{\delta (\overline{K}-%
\overline{K}^{\prime })\delta (\overline{P}-\overline{P}^{\prime }),\delta (%
\overline{S}-\overline{S}^{\prime })\}  \label{a4.56}
\end{equation}
Integral over $\overline{w}$ concludes integration over all continuous and
discrete states
\begin{equation}
\int (.)d\overline{w}=\int (.)d\overline{K}d\overline{P}+\int (.)d\overline{S%
}  \label{a4.57}
\end{equation}
Besides the following designations like Eq.(\ref{a4.29}) will be used
\[
\overline{w}=\{(\overline{K},\overline{P}),\overline{S}\},\qquad -\overline{w%
}=\{(\overline{P},\overline{K}),\overline{S}\},
\]
\begin{equation}
\delta (\overline{w}+\overline{w}^{\prime })=\{\delta (\overline{K}-%
\overline{P}^{\prime })\delta (\overline{P}-\overline{K}^{\prime }),\delta (%
\overline{S}-\overline{S}^{\prime })\}  \label{a4.58}
\end{equation}
The bound states arise at definite values of $\lambda $ only. The result
depends on the dimension $n$. At $n=1$ there is a root of $\Delta _{1}(z)$,
if $0<\lambda <4\pi m^{2}$. There is no root, if $\lambda <0\vee \lambda
>4\pi m^{2}$. At $n=2$ there is a root of $\Delta _{2}(z)$, if $0<\lambda
<8\pi m$. There is no root, if $\lambda \leq 0\vee \lambda >8\pi m$. At $%
n=3\quad \lambda $ is dimensionless quantity. There are no roots at any
value of $\lambda $ except for the case, when $\lambda $ is an infinitesimal
quantity and $\lambda >0$. If $\lambda $ depends on $\mu $ in such a way
that
\begin{equation}
\lim_{\mu \rightarrow +\infty }\left[ {\frac{\lambda }{8\pi }}\log \left(
\frac{\mu }{8\pi }+\sqrt{\frac{\mu ^{2}}{4m^{2}}+1}\right) -1\right] =+0
\label{a4.59}
\end{equation}
then the bound state described by Eq.(\ref{a4.50}) can exist. This depends
neither on $\lambda $, nor on the diverging part of Eq.(\ref{a4.43}). But
the mass $M$ of the bound state cannot be determined from Eq. (\ref{a4.43}).
It should be given independently.

The solution (\ref{a4.37}) of Eqs. (\ref{a4.34}) which satisfies the
conditions III-V has the form
\begin{equation}
D_{1}^{\prime }(t,t^{\prime };w,w^{\prime })=\int \kappa _{\overline{w}%
}^{(+)}(w)e^{i[\omega (\overline{w})-\omega (w)]t}[\kappa _{\overline{w}%
}^{(-)}(w^{\prime })+\kappa _{-\overline{w}}^{(-)}(w^{\prime })]e^{-i[\omega
(\overline{w})-\omega (w^{\prime })]t^{\prime }}d\overline{w}  \label{a4.60}
\end{equation}
Then it follows from Eqs. (\ref{a4.21}), (\ref{a4.3})
\begin{equation}
D_{1}(t,t^{\prime };w,w^{\prime })=[D_{1}^{\prime }(t,t^{\prime
};w,w^{\prime })-\delta (w-w^{\prime })-\delta (w+w^{\prime
})]e^{-i\varepsilon _{k}E({\bf k})t+i\varepsilon _{k^{\prime }}E({\bf %
k^{\prime }})t{\bf ^{\prime }}}  \label{a4.61}
\end{equation}
At $t=t^{\prime }=0$ and $\lambda =0$ the simultaneous commutation relation (%
\ref{a4.19}) takes the form
\begin{equation}
\lbrack b(K),b^{\ast }(K^{\prime })]_{-}=\delta (K-K^{\prime })+O_{2,2}
\label{a4.62}
\end{equation}
where the designations are used
\begin{equation}
O_{k,l}=\sum_{s=0}^{\infty }\int \int f_{k+s,l+s}({\cal K}^{k+s},{\cal K}%
^{\prime l+s})B_{k+s}^{\ast }({\cal K}^{k+s})B_{l+s}({\cal K}^{\prime l+s})d%
{\cal K}^{k+s}d{\cal K}^{\prime l+s})  \label{a4.63}
\end{equation}
$f_{k+s,l+s}$ is some function of arguments ${\cal K}^{k+s},{\cal K}%
^{^{\prime }l+s}$.

In the two-WL case, where Eq.(\ref{a4.62}) was obtained, the relations (\ref
{a4.62}) coincides with Eq.(\ref{a2.15}) at $\lambda =0.$ But at $\lambda
\neq 0$ the simultaneous commutation relation does not coincide with the
free simultaneous relation (\ref{a2.15}).

An attempt of using the simultaneous commutation relation (\ref{a2.15}) as
initial conditions of solution of the system (\ref{a4.34}) leads to the
result (\ref{a4.37}) with $B_{\overline{w}\overline{w}^{\prime }}$ defined
by the expression
\begin{equation}
B_{\overline{w}\overline{w}^{\prime }}=\int \overline{\kappa }_{\overline{w}%
}^{(-)}(w)[\overline{\kappa }_{\overline{w}^{\prime }}^{(+)}(w^{\prime })+%
\overline{\kappa }_{\overline{w}^{\prime }}^{(+)}(-w)]dw  \label{a4.64}
\end{equation}
In explicit form the expression (\ref{a4.64}) is written for $\overline{w}=\{%
\overline{K},\overline{P}\},\quad \overline{w}^{\prime }=\{\overline{K}%
^{\prime },\overline{P}^{\prime }\}$
\[
B_{\overline{w}\overline{w}^{\prime }}=\delta (\overline{w}-\overline{w}%
^{\prime })+\delta (\overline{w}+\overline{w}^{\prime })\frac{\lambda }{%
\left( 2\pi \right) ^{n}}\frac{\eta _{2}(\overline{w})\eta _{2}(\overline{w}%
^{\prime })\delta ({\bf \overline{s}}-{\bf \overline{s}^{\prime }})}{[\omega
(\overline{w})-\omega (\overline{w}^{\prime })\pm i0]\Delta _{+}(\overline{w}%
^{\prime })\Delta _{-}(\overline{w})}
\]
\begin{equation}
\times \{2\varepsilon _{\overline{k}}\Delta _{+}(\overline{w}^{\prime
})-2\varepsilon _{\overline{k}^{\prime }}\Delta _{+}(\overline{w}%
)-(\varepsilon _{\overline{k}^{\prime }}+\varepsilon _{\overline{k}})[\tilde{%
\Delta}_{+}(\overline{w}^{\prime })-\tilde{\Delta}_{+}(\overline{w})]\}
\label{a4.65}
\end{equation}
where
\[
\tilde{\Delta}_{\pm }(\overline{w})=1+\lambda \tilde{I}^{\pm }(\omega (%
\overline{w}),{\bf s})
\]
\begin{equation}
\tilde{I}^{(\pm )}(\omega ,{\bf s})=\frac{1}{(2\pi )^{n}}\int \frac{\eta
_{2}^{2}(w)\delta ({\bf s}-{\bf k}-{\bf p})}{\omega -\omega (w)\pm i0}dw
\label{a4.66}
\end{equation}
The functions $\eta _{2}$, $\omega (\overline{w})$ are defined by Eq.(\ref
{a4.36}). The functions $\Delta _{\pm }$ are defined by Eqs. (\ref{a4.39}), (%
\ref{a4.49}). The commutation relation is supposed to be translation
invariant. If this condition is fulfilled (condition III), then the function
$F_{1}$ determined by Eqs.(\ref{a4.3}), (\ref{a4.37}), (\ref{a4.65}) has to
depend on the difference $t-t^{\prime }$, and $B_{\overline{w}\overline{w}%
^{\prime }}$ has to contain the factor $\delta \lbrack \omega (\overline{w}%
)-\omega (\overline{w}^{\prime })]$. It is easy to verify that it is
possible in the only case, when $\lambda =0,$ i.e. for the free field.

Thus, at the quantization according to $WL$-scheme the conventional form of
simultaneous commutation relation occurs to be incompatible with the
translation invariance. It cannot be used at the nonlinear field
quantization according to $WL$-scheme.

At the quantization {\it according to} $WL${\it -scheme the simultaneous
commutation relation depends on} $\lambda $ in general.

{\it Remark}. Expression for the function $D_{1}^{\prime }$ can be taken in
the form
\begin{equation}
D_{1}^{\prime }(t,t^{\prime };w,w^{\prime })=\int \kappa _{\overline{w}%
}^{(-)}(w)e^{i[\omega (\overline{w})-\omega (w)]t}[\kappa _{\overline{w}%
}^{(+)}(w^{\prime })+\kappa _{-\overline{w}}^{(+)}(w^{\prime })]e^{-i[\omega
(\overline{w})-\omega (w^{\prime })]t^{\prime }}d\overline{w}  \label{a4.67}
\end{equation}
The expression for $D_{1}^{\prime }$ is obtained from Eq.(\ref{a4.60}) by
the substitution $\kappa ^{(+)}\rightarrow \kappa ^{(-)}$, $\kappa
^{(-)}\rightarrow \kappa ^{(+)}$ satisfies the conditions II-VI.

\section{Free Fields}

Let ${\cal H}_{2}$ be a Hilbert space of states, containing not more than
two WLs, i.e. ${\cal H}_{2}$ consists of vectors of the form (\ref{a4.17})
with $l\leq 2.$ Two-WL states $|K,P\rangle $ are not orthonormal. Indeed,
due to Eqs. (\ref{a4.3}), (\ref{a4.60}) the simultaneous commutation
relation (\ref{a4.20}) takes the form
\[
b(K)b^{\ast }(K^{\prime })=\delta (K-K^{\prime })[1-\int b^{\ast }(P)b(P)dP
\]
\begin{equation}
+\int \int D_{1}^{\prime }(K,P;K^{\prime },P^{\prime })b^{\ast
}(P)b(P^{\prime })dPdP^{\prime }]+O_{2,2}  \label{a5.1}
\end{equation}
where
\begin{equation}
D_{1}^{\prime }(K,P;K^{\prime }P^{\prime })=D_{1}^{\prime }(0,0;w,w^{\prime
})  \label{a5.2}
\end{equation}
and $O_{2,2}$ is defined by Eq. (\ref{a4.63}). Then
\begin{equation}
\langle w^{\prime }|w\rangle \equiv \langle P^{\prime },K^{\prime
}|K,P\rangle \equiv {\frac{1}{2}}\langle 0|b(P^{\prime })b(K^{\prime
})b^{\ast }(K)b^{\ast }(P)|0\rangle ={\frac{1}{2}}D_{1}^{\prime
}(K,P;K^{\prime }P^{\prime })  \label{a5.3}
\end{equation}
that is distinguished from $[\delta (w^{\prime }-w)+\delta (w+w^{\prime
})]/2 $, if $\lambda \neq 0$.

Let us introduce orthonormal states $|\overline{w}\rangle _{+}$, $|\overline{%
w}\rangle _{-}$, defining them by relations
\begin{equation}
|\overline{w}\rangle _{\alpha }=\{|\overline{K},\overline{P}\rangle _{\alpha
},|\overline{S}\rangle _{\alpha }\},\qquad \alpha =\pm  \label{a5.4}
\end{equation}
\begin{equation}
|\overline{K},\overline{P}\rangle _{\alpha }=\int \!\!\int \overline{\kappa }%
_{\overline{K},\overline{P}}^{(\alpha )}(K,P)|K,P\rangle dKdP=\int \kappa _{%
\overline{w}}^{(\alpha )}(w)|w\rangle dw,\qquad \alpha =\pm  \label{a5.5}
\end{equation}
\begin{equation}
|S\rangle _{\alpha }=\int \!\!\int \overline{\kappa }_{\overline{S}%
}(K,P)|K,P\rangle dKdP,\qquad \alpha =\pm  \label{a5.6}
\end{equation}
\noindent The wave functions $f(K,P),$
\[
\psi _{(+)}(\overline{w})=\{\psi _{(+)}(\overline{K},\overline{P}),\psi
(S)\},\quad \psi _{(-)}(\overline{w})=\{\psi _{(-)}(\overline{K},\overline{P}%
),\psi (S)\}
\]
of the state $\Phi _{2}$ in these three different representations are
defined by the relations
\begin{equation}
\Phi _{2}=\int f(w)|w\rangle dw=\int \psi _{(+)}(\overline{w})|\overline{w}%
\rangle _{+}d\overline{w}=\int \psi _{(-)}(\overline{w})|\overline{w}\rangle
_{-}d\overline{w}.  \label{a5.7}
\end{equation}
They are connected by means of relations
\[
\psi _{(\pm )}(\overline{w})=\int \kappa _{\overline{w}}^{(\mp )}(w)f(w)d%
\overline{w},
\]
\begin{equation}
f(w)=\int \overline{\kappa }_{\overline{w}}^{(\pm )}(w)\psi _{(\pm )}(%
\overline{w})d\overline{w},  \label{a5.8}
\end{equation}
\[
\psi _{(\alpha )}(\overline{w})=\int \Omega ^{(-\alpha )}(w)\psi _{(-\alpha
)}(\overline{w}^{\prime })d\overline{w}^{\prime },\qquad \alpha =\pm ,
\]
where
\[
\Omega ^{(\pm )}(\overline{w},\overline{w}^{\prime })=\int \overline{\kappa }%
_{\overline{w}}^{(\pm )}(w)\kappa _{\overline{w}^{\prime }}^{(\pm )}(w)dw,
\]
\[
\Omega ^{(\pm )}(\overline{K},\overline{P};\overline{K}^{\prime }\overline{P}%
^{\prime })=\delta (\overline{w}-\overline{w}^{\prime })\pm \frac{i\lambda }{%
(2\pi )^{n-1}}\frac{\varepsilon _{\overline{k}}\eta _{2}(\overline{w})\eta
_{2}(\overline{w}^{\prime })}{\Delta _{\pm }(\overline{w})}\delta \lbrack
\omega (\overline{w})-\omega (\overline{w}^{\prime })]
\]
\begin{equation}
\times \delta ({\bf \overline{k}}+{\bf \overline{p}}-{\bf \overline{k}%
^{\prime }}-{\bf \overline{p}^{\prime }}),  \label{a5.9}
\end{equation}
\[
\Omega ^{(\pm )}(\overline{S};\overline{S}^{\prime })=\delta (\overline{S}-%
\overline{S}^{\prime }),
\]
\[
\Omega ^{(\pm )}(\overline{K},\overline{P};\overline{S}^{\prime })=\Omega
^{(\pm )}(\overline{S};\overline{K},\overline{P})=0.
\]
Here the upper signs, or the lower ones are taken simultaneously.

One can see from Eq. (\ref{a5.8}) that a symmetrization of anyone of
functions $f(w),$ $\psi _{(\pm )}(\overline{w}),$ $\psi _{(-)}(\overline{w})$
leads to symmetrization of remaining functions, i.e. a fulfillment of anyone
of equalities
\[
f(w)=f(-w),\qquad \psi _{(+)}(\overline{w})=\psi _{(+)}(-\overline{w}),
\]
\begin{equation}
\psi _{(-)}(\overline{w})=\psi _{(-)}(-\overline{w})  \label{a5.10}
\end{equation}
leads to fulfillment of two others.

The wave function of identical WLs has to be symmetric. In QFT the symmetric
wave functions arise as a result of all creation operators commutativity (%
\ref{a2.15}). It leads to identification of the states $|K,P\rangle $ and $%
|P,K\rangle $ in the form
\begin{equation}
|K,P\rangle -|P,K\rangle ={\frac{1}{\sqrt{2}}}[b^{\ast }(K)b^{\ast
}(P)-b^{\ast }(P)b^{\ast }(K)]|0\rangle =0  \label{a5.11}
\end{equation}
But it is possible only for a free field. Indeed, convoluting Eq. (\ref
{a5.11}) with the state vector $\langle P^{\prime },K^{\prime }|$, one
obtains by means of Eqs.(\ref{a5.3}) and (\ref{a4.60})
\begin{equation}
D_{1}^{\prime }(w^{\prime },w)-D_{1}^{\prime }(w^{\prime },-w)=0
\label{a5.12}
\end{equation}
For $\varepsilon _{k}=-\varepsilon _{p}$ Eq. (\ref{a5.12}) is fulfilled only
at $\lambda =0.$

Let us introduce operators of free field $b_{0}(K,t)$, $b_{1}(S,t)$, $%
b_{0}^{\ast }(K,t)$, $b_{1}^{\ast }(S,t)$, defining them by relations
\begin{equation}
b_{0}(K,t)|0\rangle =b_{1}(K,t)|0\rangle =0  \label{a5.13}
\end{equation}
\begin{equation}
b_{0}^{\ast }(K,t)|0\rangle =b^{\ast }(K,t)|0\rangle  \label{a5.14}
\end{equation}
\begin{equation}
b_{0}^{\ast }(\overline{K},t)b_{0}^{\ast }(\overline{P},t)|0\rangle =\int
\int \overline{\kappa }_{\overline{K},\overline{P}}^{(+)}(K,P)b^{\ast
}(K,t)b^{\ast }(P,t)|0\rangle dKdP  \label{a5.15}
\end{equation}
\begin{equation}
b_{1}^{\ast }(\overline{S},t)|0\rangle =\int \int \overline{\kappa }_{%
\overline{S}}(K,P)b^{\ast }(K,t)b^{\ast }(P,t)|0\rangle dKdP  \label{a5.16}
\end{equation}
The solution of Eqs. (\ref{a5.13})-(\ref{a5.16}) can be represented in the
form
\begin{equation}
b_{0}(K,t)=b(K,t)+\int \int \int \overline{\mu }_{K,P}^{(-)}(K^{\prime
},P^{\prime })b^{\ast }(P,t)b(P^{\prime },t)b(K^{\prime },t)dK^{\prime
}dP^{\prime }dP+O_{2,3}  \label{a5.17}
\end{equation}
\begin{equation}
b_{1}(S,t)=\int \int \overline{\mu }_{S}(K^{\prime },P^{\prime })b(P^{\prime
},t)b(K^{\prime },t)dK^{\prime }dP^{\prime }+O_{1,3}  \label{a5.18}
\end{equation}
where
\[
\overline{\mu }_{\overline{K},\overline{P}}^{(\pm )}(K,P)=\overline{\kappa }%
_{\overline{K},\overline{P}}^{(\pm )}(K,P)-\delta (\overline{K}-K)\delta (%
\overline{P}-P)=
\]
\begin{equation}
-\frac{\lambda }{(2\pi )^{n}}\frac{\eta _{1}(\overline{w})\eta _{2}(w)\delta
({\bf s}-{\bf \overline{s}})}{[\omega (\overline{w})-\omega (w)\pm i0]\Delta
_{\pm }(\overline{w})}  \label{a5.19}
\end{equation}
\begin{equation}
\overline{\mu }_{\overline{S}}(K,P)={\frac{1}{\sqrt{2}}}\overline{\kappa }_{%
\overline{S}}(K,P)=\frac{\varepsilon _{s}\eta _{2}(w)\delta ({\bf s}-{\bf
\overline{s}})}{\sqrt{(2\pi )^{n}B}\sqrt{\beta _{{\rm M}}({\bf s})}[\omega (%
\overline{S})-\omega (w)]}  \label{a5.20}
\end{equation}
\noindent where $B$ is defined by Eq.(\ref{a4.51})

The reciprocal relation has the form
\begin{eqnarray}
b(K,t) &=&b_{0}(K,t)+\int \int \mu _{K^{\prime },P^{\prime
}}^{(+)}(K,P)b_{0}^{\ast }(P,t)b_{0}(P^{\prime },t)b_{0}(K^{\prime
},t)dK^{\prime }dP^{\prime }dP  \nonumber \\
&&+\int \int \mu _{S}(K,P)b_{0}^{\ast }(P,t)b_{1}(K,t)dSdP+O_{2,3}
\label{a5.21}
\end{eqnarray}
where
\begin{equation}
\mu _{\overline{K},\overline{P}}^{(\pm )}(K,P)=\kappa _{\overline{K},%
\overline{P}}^{(\pm )}(K,P)-\delta (\overline{K}-K)\delta (\overline{P}-P)
\label{a5.22}
\end{equation}
\begin{equation}
\mu _{\overline{S}}(K,P)=\sqrt{2}\kappa _{\overline{S}}(K,P)  \label{a5.23}
\end{equation}
the relations for $b_{0}^{\ast }$, $b_{1}^{\ast }$, $b$ are obtained from
Eqs.(\ref{a5.17}), (\ref{a5.18}), (\ref{a5.21}) by means of a Hermitian
conjugation.

Let the commutation relation have the form (\ref{a5.1}), where $%
D_{1}^{\prime }$ is defined by Eq. (\ref{a4.60}). Then a calculation gives
\begin{equation}
\lbrack b_{0}(K,t),b_{0}^{\ast }(K^{\prime },t)]_{-}=\delta (K-K^{\prime
})+O_{2,2}  \label{a5.24}
\end{equation}
\begin{equation}
b_{0}(K,t)b_{1}^{\ast }(K^{\prime },t)=O_{2,1}  \label{a5.25}
\end{equation}
\begin{equation}
b_{1}(K,t)b_{0}^{\ast }(K^{\prime },t)=O_{1,2}  \label{a5.26}
\end{equation}
\begin{equation}
b_{1}(K,t)b_{1}^{\ast }(K^{\prime },t)=\delta (K-K^{\prime })+O_{2,2}
\label{a5.27}
\end{equation}
Differentiating Eqs.(\ref{a5.17}), (\ref{a5.18}) with respect to $t$ and
using Eqs.(\ref{a4.3}), (\ref{a4.4}), (\ref{a5.1}), one obtains
\begin{equation}
\dot{b}_{A}(K,t)=-i\varepsilon _{k}E_{A}({\bf k})b_{A}(K,t)+O_{2-A,3},\qquad
A=0,1  \label{a5.28}
\end{equation}
\begin{equation}
E_{A}(k)=|\sqrt{m_{A}^{2}+{\bf k}^{2}}|,\qquad m_{0}=m,\qquad m_{1}=M,\qquad
A=0,1  \label{a5.29}
\end{equation}
Let us introduce the fields $\varphi _{A}(x)=\varphi _{A}({\bf x},t)$, $%
A=0,1 $, defining them by relations of the type (\ref{a2.4}), (\ref{a2.5})
\begin{equation}
\varphi _{A}({\bf x},t)=(2\pi )^{-n/2}\int e^{i{\bf kx}}\frac{b_{A}(K,t)}{%
\sqrt{\beta _{A}({\bf k})}}dK,\qquad \beta _{A}({\bf k})=2E_{A}({\bf k}%
),\qquad A=0,1  \label{a5.30}
\end{equation}
\begin{equation}
\dot{\varphi}_{A}({\bf x},t)=\frac{\partial \varphi _{A}}{\partial t}%
=-i(2\pi )^{-n/2}\int {\frac{\varepsilon _{k}}{2}}\sqrt{\beta _{A}({\bf k})}%
e^{i{\bf kx}}b_{A}(K,t)dK,\qquad A=0,1  \label{a5.31}
\end{equation}
Then according to Eq.(\ref{a5.23}) the fields $\varphi _{A}(x)$ satisfy the
dynamic equations
\begin{equation}
(\partial _{i}\partial ^{i}+m_{A}^{2})\varphi _{A}=O_{2-A,3},\qquad A=0,1
\label{a5.32}
\end{equation}
In other words, inside the Hilbert space ${\cal H}_{2}$ the fields $\varphi
_{A}(x)$, $A=0,1$ are free non-interacting fields of the mass $m_{A}$.

Using the transformation properties of $\varphi $ and relations (\ref{a5.17}%
), (\ref{a5.18}), one can show the fields $\varphi _{0},\varphi _{1}$ are
scalars. Thus, nonlinear field $\varphi $ can be described by means of two
free scalar fields $\varphi _{0},\varphi _{1}$ to within $O_{2,3}$.

In the case of the commutation relations (\ref{a5.24}) -- (\ref{a5.27}) the
Hamiltonian $H=-p_{0}$ and the canonical momentum $\pi _{\alpha }=-p_{\alpha
}$ defined by Eq. (\ref{a3.11}) have the following form
\begin{equation}
H=\sum_{A=0,1}^{{}}\int \varepsilon _{k}E_{A}({\bf k})b_{A}^{\ast
}(K)b_{A}(K)dK+O_{3,3}  \label{a5.33}
\end{equation}
\begin{equation}
{\bf \pi }=\sum_{A=0,1}\int {\bf k}b_{A}^{\ast }(K)b_{A}(K)dK+O_{3,3}
\label{a5.34}
\end{equation}
Vectors $|0\rangle ,\quad |K\rangle $,\quad $|K,P\rangle _{+}+|P,K\rangle
_{+}$,\quad $|K,P\rangle _{-}+|P,K\rangle _{-}$,\quad $b_{1}^{\ast
}(S)|0\rangle $ are eigenvectors of operators $H$, $\bpi$ with eigenvalues $%
\{0,{\bf 0}\}$, $\{\varepsilon _{k}E({\bf k}),{\bf k}\}$, $\{\omega (K,P)+i0,%
{\bf k}+{\bf p}\}$, $\{\omega (K,P)-i0,{\bf k}+{\bf p}\}$, $\{\varepsilon
_{s}E_{M}({\bf k}),{\bf s}\}$ respectively. All of them describe stationary
states. Vectors $|K,P\rangle _{+}$, $|P,K\rangle _{+}$, $|K,P\rangle
_{-},|P,K\rangle _{-}$ are not stationary states at $K\neq P$, generally.

The commutation relations (\ref{a5.24}) -- (\ref{a5.27}) permit only to
commutate the creation operator $b^{\ast }$ with the annihilation operator $%
b $. They do not permit to commutate two creation operators $b^{\ast }$. For
this reason the vector $|K,P\rangle _{+}=2^{-1/2}b_{0}^{\ast }(K)b_{0}^{\ast
}(P)|0\rangle $ does not connect with the vector $|P,K\rangle
_{+}=2^{-1/2}b_{0}^{\ast }(P)b_{0}^{\ast }(K)|0\rangle $, if $K\neq P$.

Consideration of the WL identity is realized by the symmetric wave functions
(\ref{a5.8}). Using the condition (\ref{a5.11}) is impossible, as we have
seen. But the states $|K,P\rangle _{+}$ and $|P,K\rangle _{+}$ can be
identified , as far as according to Eq. (\ref{a5.24}) the vector $%
|K,P\rangle _{+}-|P,K\rangle _{+}$ is orthogonal to all vectors of ${\cal H}%
_{2}$.
\begin{equation}
\langle P^{\prime },K^{\prime }|K,P\rangle _{+}-\langle P^{\prime
},K^{\prime }|P,K\rangle _{+}=0  \label{a5.35}
\end{equation}
Let us identify the vector $|K,P\rangle _{+}-|P,K\rangle _{+}$ with the zero
vector. It is equivalent to the commutation relations
\begin{equation}
\lbrack b_{0}^{\ast }(K),b_{0}^{\ast }(K^{\prime })]_{-}=O_{3,1}
\label{a5.36}
\end{equation}
\begin{equation}
\lbrack b^{\ast }(K),b^{\ast }(P)]_{-}=\frac{i\lambda }{(2\pi )^{n}}\int
\frac{(\varepsilon _{k}-\varepsilon _{p})\eta _{2}(w)\eta _{2}(\overline{w}%
)\delta ({\bf s}-{\bf \overline{s}})}{[\omega (\overline{w})-\omega
(w)-i0]\Delta _{-}(\overline{w})}b_{0}^{\ast }(\overline{K})b_{0}^{\ast }(%
\overline{P})d\overline{K}d\overline{P}+O_{3,1}  \label{a5.37}
\end{equation}
If the condition (\ref{a5.36}) takes place, then the Hilbert space ${\cal H}%
_{2}$ turns to the Hilbert space ${\cal H}_{2,s}$ of identical WLs.

Let us introduce operators $S^{(+)}$ and $S^{(-)}$ by means of relations
\begin{equation}
S^{(\pm )}=1+\int \int \sigma ^{(\pm )}(w,w^{\prime })b_{0}^{\ast
}(P)b_{0}^{\ast }(K)b_{0}(K^{\prime })b_{0}(P^{\prime })dwdw^{\prime
}+O_{3,3}  \label{a5.38}
\end{equation}
where
\[
\sigma ^{(\pm )}(w,w^{\prime })={\frac{1}{2}}\{\Omega ^{(\pm
)}(K,P;K^{\prime },P^{\prime })-\delta (w-w^{\prime })\}+(K\leftrightarrow
P)
\]
\begin{equation}
\pm \frac{i\lambda }{(2\pi )^{n-1}}\frac{(\varepsilon _{k}+\varepsilon
_{p})\eta _{2}(w)\eta _{2}(w^{\prime })}{2\Delta _{\pm }(w)}\delta \lbrack
\omega (w)-\omega (w^{\prime })]\delta ({\bf k}+{\bf p}-{\bf k^{\prime }}-%
{\bf p^{\prime }})  \label{a5.39}
\end{equation}
where $\Omega ^{(\pm )}$ is defined by Eqs. (\ref{a5.9}), and $%
(K\leftrightarrow P)$ means the term, obtained from the preceding one by the
substitution $K\leftrightarrow P$.

One can verify that operators $S^{(+)}$ and $S^{(-)}$ are unitary in ${\cal H%
}_{2,s}$
\begin{equation}
(S^{(+)})^{\ast }=S^{(-)}+O_{3,3},\qquad
S^{(+)}S^{(-)}=S^{(-)}S^{(+)}=1+O_{3,3}  \label{a5.40}
\end{equation}
\noindent and commutate with operators $H$ and $\bpi$
\begin{equation}
\lbrack S^{(\pm )},H]_{-}=O_{3,3},\qquad \lbrack S^{(\pm )},\bpi]_{-}=O_{3,3}
\label{a5.41}
\end{equation}
Let us introduce operators
\begin{equation}
c_{0}^{\ast }(K,t)=S^{(-)}b_{0}^{\ast }(K,t)S^{(+)},\qquad
c_{0}(K,t)=S^{(-)}b_{0}(K,t)S^{(+)}  \label{a5.42}
\end{equation}
From unitary property of $S^{(+)}$ and the relation
\begin{equation}
c_{1}(K,t)=S^{(-)}b_{1}(K,t)S^{(+)}=b_{1}(K,t)+O_{1,3}.  \label{a5.43}
\end{equation}
it follows that operators $c_{0}$, $c_{1}$ satisfy the same commutation
relations (\ref{a5.24})-(\ref{a5.27}), (\ref{a5.36}), as operators $b_{0}$, $%
b_{1}$. From relations (\ref{a5.41}) it follows that operators $c_{0}(K,t)$,
$c_{1}(K,t)$ satisfy the same equations (\ref{a5.28}), as operators $%
b_{0}(K,t)$, $b_{1}(K,t)$ do.

In other words, the operators $c_{0}(K,t)$, $b_{0}(K,t)$ describe the free
fields and satisfy the same commutation relation. Operators $b_{0}$, $c_{0}$
have many properties common with $in$- and $out$-operators in QFT [9]. But
there are differences. For instance, the simultaneous commutation relations
are the same for operators $b_{{\rm in}}, b_{{\rm out}}$ and $b$, whereas
they are different for $b_{0}$ and $b$. For this reason we do not identify
operators $b_{0}$, $c_{0}$ with $b_{{\rm in}}$ and $b_{{\rm out}}$.

From Eqs.(\ref{a5.5}), (\ref{a5.6}), (\ref{a5.9}), (\ref{a5.38}), (\ref
{a5.39}) it follows
\begin{equation}
|K,P\rangle _{-}=S^{(-)}|K,P\rangle _{+},\qquad |K,P\rangle
_{+}=S^{(+)}|K,P\rangle _{-}  \label{a5.44}
\end{equation}
According to Eqs. (\ref{a5.42}), (\ref{a5.15}) it can be written in the form
\begin{equation}
{\frac{1}{\sqrt{2}}}c_{0}^{\ast }(K)c_{0}^{\ast }(P)|0\rangle
=S^{(-)}|K,P\rangle _{+}=|K,P\rangle _{-}  \label{a5.45}
\end{equation}
The operator $S^{(-)}$ is usually referred to as the scattering matrix or $S$%
-matrix. Further it will be shown that, indeed, the operator $S^{(-)}$
describes the SWL scattering. But it is impossible to make {\it in the scope
of the quantization scheme.} In addition the operator, describing the
spatial SWL distribution, has to be introduced, for instance, the SWL
current density.

\section{The Scattering Problem}

At the conventional approach [1] to the relativistic scattering problem the
interaction cut-off at $t \rightarrow \infty $ is used. In other words, at
the scattering problem statement one uses two Hamiltonians: the Hamiltonian $%
H$ defined by Eq.(\ref{a5.3}) and unperturbed Hamiltonian $H_{0}$. But the
real interaction cannot be cut off. There is only one Hamiltonian $H$, and
{\it the scattering problem should be stated in terms of only this
Hamiltonian} $H$.

Let us state the scattering problem as follows. Let the state $\Phi _{2}\in
{\cal H}_{2,s}$ describe two wave packets of particles, moving one through
another at the time moment $t=0$ at the origin of the coordinate system.
Each of two wave packets is supposed to be almost monomomentum and to have
the size $a\gg m^{-1}$, where $m^{-1}$ is the Compton wave length of the
particle. It is necessary, for the spread of the wave packets could be
neglected. Let the particles of the wave packets be recorded by detectors of
the size $L\gg a$. The detectors are placed at such a large distance $d\gg L$
from the frame origin, that the detector angular size $\Omega =L/d\ll 1.$ In
this case the particles of one of the wave packets are recorded by a
detector $\alpha _{1}$ and those of the other one are recorded by a detector
$\alpha _{2}$. The scattered particle, (if there are such ones) are recorded
by detectors $\beta _{1},\beta _{2},\ldots $

Let $\Phi ^\prime _2\in {\cal H}_{2,s}$ be another state, which is
distinguished from $\Phi _{2}$ only by some displacement of the second wave
packet. It is displaced at the distance $l$, $(L \gg l \gg a)$ in such a way
that it does not go through the first wave packet. In this case the
particles of the first wave packet are recorded by the same detector $\alpha
_{1}$, those of the second one are recorded by the detector $\alpha _{2}$.
There are no scattered particles in this case. Thus, one can investigate the
scattering, comparing the particle densities in the states $\Phi _{1}$ and $%
\Phi _{2}^{\prime}$.

Let $d\nu $ be the number of particles scattered in the direction ${\bf l}$
inside the solid angle $d\omega $. Let $n_{1}^{i}({\bf x},t),n_{2}^{i}({\bf x%
},t)$ be the flux densities of particles in the first wave packet and in the
second one respectively. Then the section $d\sigma $ of the scattering in
the direction ${\bf l}$ is described by the relation
\begin{equation}
d\nu =Jd\sigma  \label{a6.1}
\end{equation}
\begin{equation}
J=\int \int \sqrt{%
(n_{1}^{i},n_{2i})^{2}-(n_{1}^{i},n_{1i})^{2}(n_{2}^{k},n_{2k})^{2}}d{\bf x}%
dt  \label{a6.2}
\end{equation}
Here the invariant integral $J$ describes a degree of overlapping of the
wave packets. In particular, in the coordinate frame, where the first wave
packet is at rest and the second one moves with the velocity ${\bf v}_{{\rm %
rel}}$ , i.e.
\begin{equation}
n_{1}^{0}=n_{1},\quad {\bf n}_{1}=0,\qquad n_{2}^{0}=n_{2},\quad {\bf n}%
_{2}=n_{2}{\bf v}_{{\rm rel}}  \label{a6.3}
\end{equation}
Eq. (\ref{a6.1}) transforms into well known expression

\begin{equation}
d\nu =d\sigma \int n_{1}n_{2}\left| {\bf v}_{{\rm rel}}\right| d{\bf x}dt.
\label{a6.4}
\end{equation}
Let us choose the state $\Phi _{2}$ in the form

\begin{equation}
\Phi _{2}=\int f_{W}(K,P)|K,P\rangle dKdP  \label{a6.5}
\end{equation}
where
\begin{equation}
f_{W}(K,P)={\frac{1}{2}}F(K-R)F(P-Q)e^{-i{\bf ky}-i{\bf pz}%
}+(K\leftrightarrow P)  \label{a6.6}
\end{equation}
Here and further $(K\leftrightarrow P)$ means the term obtained from the
preceding one by the substitution $K\leftrightarrow P$. $W=\{R,Q,{\bf y},%
{\bf z}\}$, $R=\{\varepsilon _{r},{\bf r}\}$, $Q=\{\varepsilon _{q},{\bf q}%
\} $ is a set of parameters describing the wave function $\psi _{W}$.
\begin{equation}
F(K-R)=A\delta _{\varepsilon _{k},\varepsilon _{r}}\exp \{-{\frac{1}{2}}(%
{\bf k}-{\bf r})^{2}a^{2}\}  \label{a6.7}
\end{equation}
is a real function which is distinguished from zero essentially only in the
region
\begin{equation}
|{\bf k}-{\bf r}|<a^{-1}\ll m  \label{a6.8}
\end{equation}
$A$ is a normalization constant.

Remaining in the scope of the quantization scheme one cannot understand the
meaning of parameters $W$ of the wave function $\psi _{W}$. Only passing to
the quantization model and introducing operators of physical quantities, on
can understand what the parameters $W$ mean. But the physical quantities can
be introduced in different ways.

The charge density $j^{0}({\bf x},t)$ can be introduced at least by two
ways. The first way
\begin{equation}
j^{0}({\bf x},t)=e^{iHt-i\bpi {\bf x}}\frac{e}{(2\pi )^{n}}\int \frac{%
\varepsilon _{k}}{2}\sqrt{\beta ({\bf k})/\beta ({\bf k^{\prime }})}b^{\ast
}(K)b(K^{\prime })dKdK^{\prime }e^{-iHt+i\bpi {\bf x}}+(\hbox{h.c.})
\label{a6.9}
\end{equation}
The second one
\begin{eqnarray}
j^{0}({\bf x},t) &=&e^{iHt-i\bpi {\bf x}}\int \frac{\varepsilon _{k}}{2(2\pi
)^{n}}\sum_{A=0,1}e_{A}\sqrt{\beta ({\bf k})/\beta ({\bf k^{\prime }})}
\nonumber \\
&&\times b_{A}^{\ast }(K)b_{A}(K^{\prime })dKdK^{\prime }e^{-iHt+i\bpi {\bf
x}}+(\hbox{h.c.})  \label{a6.10}
\end{eqnarray}
Here (h.c.) means an addition of Hermitian conjugate expression. $e_{0}=e$, $%
e_{1}=2e$.

Eq.(\ref{a6.9}) corresponds to expression (\ref{a3.10}) for the nonlinear
field $\varphi $. Eq.(\ref{a6.10}) corresponds to the same expression (\ref
{a3.10}) for linear noninteracting fields $\varphi _{0},\varphi _{1}$. The
total charge $Q$ is the same in both cases.
\begin{equation}
Q=e\int \varepsilon _{k}b^{\ast }((K)b(K)dK=\sum_{A=0,1}^{{}}e\int
\varepsilon _{k}b_{A}^{\ast }((K)b_{A}(K)dK+O_{3,3}  \label{a6.11}
\end{equation}
In the second model of quantization $j^{0}$ is defined by Eq. (\ref{a6.10}),
the state (\ref{a6.5}), (\ref{a6.6}) describes two wave packets of the size $%
\cong a$. They move with velocities
\begin{equation}
{\bf v}_{1}=\frac{\varepsilon _{r}{\bf r}}{E({\bf r})},\qquad {\bf v}_{2}=%
\frac{\varepsilon _{q}{\bf q}}{E({\bf q})}  \label{a6.12}
\end{equation}
and are placed at the moment $t=0$ at the points ${\bf y}$ and ${\bf z}$
respectively. There are {\it no scattered particles} in the second
quantization model.

In the first quantization model, where $j^{0}$ is defined by Eq. (\ref{a6.9}%
), the state (\ref{a6.5}), (\ref{a6.6}) describes the same wave packets as
in the second model. But in this case there are {\it "scattered charges", if
the wave packets overlap.}

Let us consider the first model in details. Averaging the operator (\ref
{a6.9}) over $\Phi _{2}$ one obtains the following expression
\[
\langle j^{0}({\bf x},t)\rangle _{\Phi _{2}}=(\Phi _{2},j^{0}({\bf x},t)\Phi
_{2})
\]
\begin{equation}
=\frac{e}{(2\pi )^{n}}\sum\limits_{\varepsilon _{k^{\prime }},\varepsilon
_{k^{\prime \prime }}=\pm 1}\int \varepsilon _{k^{\prime }}\Psi ^{(+)}({\bf x%
},t;\varepsilon _{k^{\prime }},P)\Psi ^{(-)}({\bf x},t;\varepsilon
_{k^{\prime \prime }},P)dP+(\hbox{c.c.})  \label{a6.13}
\end{equation}
where
\begin{equation}
\Psi ^{(\pm )}({\bf x},t;\varepsilon _{k^{\prime \prime }},P)=\Psi _{{\rm fr}%
}^{(\pm )}({\bf x},t;\varepsilon _{k^{\prime \prime }},P)+\Psi _{{\rm sc}%
}^{(\pm )}({\bf x},t;\varepsilon _{k^{\prime \prime }},P)  \label{a6.14}
\end{equation}
\begin{equation}
\Psi _{{\rm fr}}^{(\pm )}({\bf x},t;\varepsilon _{k^{\prime \prime }},P)={%
\frac{1}{2}}\int F(K^{\prime \prime }-R)F(P^{\prime \prime }-Q)e^{\pm i\zeta
_{{\rm fr}}(K^{\prime \prime },P^{\prime \prime })}d{\bf k}^{\prime \prime
}+(W\rightarrow -W)  \label{a6.15}
\end{equation}
\[
\Psi _{{\rm sc}}^{(\pm )}({\bf x},t;\varepsilon _{k^{\prime \prime
}},P^{\prime \prime })={\frac{1}{2}}\int F(K-R)F(P-Q)\mu _{K,P}^{(\mp
)}(K^{\prime \prime },P^{\prime \prime })
\]
\begin{equation}
\times e^{\pm i\zeta _{{\rm fr}}(K,P)}dKdPd{\bf k}^{\prime \prime
}+(W\rightarrow -W)  \label{a6.16}
\end{equation}
Here and further $(W\rightarrow -W)$ means the term obtained from the
preceding one by means of transposition $(R\leftrightarrow Q$, ${\bf y}%
\leftrightarrow {\bf z})$.
\begin{equation}
\zeta _{{\rm fr}}(K,P)=\omega (K,P)t-{\bf k}({\bf x}-{\bf y})-{\bf p}({\bf x}%
-{\bf z})  \label{a6.17}
\end{equation}
where $\omega (K,P)$ is defined by Eq. (\ref{a4.36}), or (\ref{a4.45}).

The functions $\mu ^{(\pm )}$ are defined by Eqs.(\ref{a4.48}), (\ref{a5.22}%
). Obtaining the expression (\ref{a6.13}), it was taken into account that
the factor $\sqrt{\beta ({\bf k})/\beta ({\bf k}^{\prime })}$ in Eq.(\ref
{a6.9}) changes slowly in the region of unvanishing contribution. $\Psi
^{(-)}({\bf x},t;\varepsilon _{k^{\prime \prime }},P)$ can be treated as a
wave function in the mixed momentum-coordinate representation. $\Psi ^{(+)}$
is a complex conjugate function. $\Psi _{fr}^{(-)},\Psi _{sc}^{(-)}$ are
responsible respectively for the free motion and for the scattering.

Using expressions (\ref{a4.48}), (\ref{a5.22}) for $\mu ^{(\pm )}$, Eq. (\ref
{a6.16}) can be represented in the form
\[
\Psi _{{\rm sc}}^{(\pm )}({\bf x},t;\varepsilon _{k^{\prime \prime
}},P^{\prime \prime })=\pm \frac{i\lambda }{(2\pi )^{n}}\int \int dKdP\frac{%
\varepsilon _{k^{\prime \prime }}\eta _{2}({\bf k}+{\bf p}-{\bf p}^{\prime
\prime },{\bf p})\eta _{2}(K,P)}{\Delta _{\pm }(K,P)}
\]
\begin{equation}
\times F(K-R)F(P-Q)\int\limits_{0}^{\infty }d\alpha e^{\pm i\zeta _{{\rm sc}%
}(K,P;\varepsilon _{k^{\prime \prime }},P^{\prime \prime },\alpha
)}+(W\rightarrow -W)  \label{a6.18}
\end{equation}
\begin{equation}
\zeta _{{\rm sc}}(K,P;\varepsilon _{k^{\prime \prime }},P^{\prime \prime
},\alpha )=\zeta _{{\rm fr}}(K,P)+\alpha \lbrack \omega (K,P)-\varepsilon
_{k^{\prime \prime }}E({\bf k}+{\bf p}-{\bf p}^{\prime \prime })-\varepsilon
_{p^{\prime \prime }}E({\bf p}^{\prime \prime })]  \label{a6.19}
\end{equation}
The wave functions (\ref{a6.15}), (\ref{a6.18}) are integrals of quickly
oscillating functions. They are distinguished from zero essentially only for
those values of parameters ${\bf x},t,W$, for which the phases $\zeta _{{\rm %
fr}}$ and $\zeta _{{\rm sc}}$ are stationary (i.e. the phases have extrema
for all arguments which are integrated over). The phase $\zeta _{{\rm fr}}$
is stationary, if the following conditions are fulfilled
\begin{equation}
{\bf x}={\bf y}+{\bf v}_{1}t,\qquad {\bf x}={\bf z}+{\bf v}_{2}t
\label{a6.20}
\end{equation}
where ${\bf v}_{1}$ and ${\bf v}_{2}$ are defined by Eq. (\ref{a6.12}).

Equations (\ref{a6.19}) describe the free motion of the wave packet centres.
In their vicinity the $\Psi ^{(\pm )}_{fr}$ is distinguished essentially
from zero.

The phase $\zeta _{sc}$ is stationary, if the following relations are
fulfilled
\begin{equation}
{\bf x}={\bf X}+\frac{\varepsilon _{k^{\prime \prime }}({\bf r}+{\bf q}-{\bf %
p}^{\prime \prime })}{E({\bf r}+{\bf q}-{\bf p}^{\prime \prime })}%
(t-T),\qquad t>T  \label{a6.21}
\end{equation}
\begin{equation}
{\bf X}={\bf y}+{\bf v}_{1}T={\bf z}+{\bf v}_{2}T  \label{a6.22}
\end{equation}
\begin{equation}
\omega (R,Q)-\varepsilon _{k^{\prime \prime }}E({\bf r}+{\bf q}-{\bf p}%
^{\prime \prime })-\varepsilon _{p^{\prime \prime }}E({\bf p^{\prime \prime }%
})=0  \label{a6.23}
\end{equation}
Conditions (\ref{a6.21}), (\ref{a6.22}) can be fulfilled only in that case,
when the wave packets, moving according to Eq. (\ref{a6.20}), pass one
through another, and their centres coincide at the time moment $T$ at the
point ${\bf X}$. If it takes place, then the ''scattered charges'' arise.
They move with the velocity
\begin{equation}
{\bf v}_{{\rm sc}}=\frac{\varepsilon _{k^{\prime \prime }}({\bf r}+{\bf q}-%
{\bf p}^{\prime \prime })}{E(r+q-p^{\prime \prime })}  \label{a6.24}
\end{equation}
in the direction from the collision point ${\bf X}$.

The admissible values ${\bf p}^{\prime \prime }$ are determined from Eq.(\ref
{a6.23}). For $\varepsilon _{r}=\varepsilon _{q}$ Eq. (\ref{a6.23}) is the
energy conservation law. In this case the scattering is elastic. Indeed, in
the coordinate system, where ${\bf q}+{\bf r}=0,$ it follows from Eq. (\ref
{a6.22})
\begin{equation}
\varepsilon _{k^{\prime \prime }}=\varepsilon _{p^{\prime \prime
}}=\varepsilon _{r}=\varepsilon _{q},\qquad \left| {\bf p}^{\prime \prime
}\right| =\left| {\bf r}\right| =\left| {\bf q}\right|  \label{a6.25}
\end{equation}
\begin{equation}
{\bf v}_{{\rm sc}}=\varepsilon _{p^{\prime \prime }}\frac{{\bf p}^{\prime
\prime }}{E({\bf p}^{\prime \prime })}\qquad \left| {\bf v}_{{\rm sc}%
}\right| =\left| {\bf v}_{1}\right| =\left| {\bf v}_{2}\right|  \label{a6.26}
\end{equation}
In the case $\varepsilon _{r}=-\varepsilon _{q}$ , when a particle collides
with an antiparticle, the scattering is absent entirely. Indeed, $\langle
j^{0}({\bf x},t)\rangle _{\Phi _{2}}$ can be separated into three parts
\begin{equation}
\langle j^{0}({\bf x},t)\rangle _{\Phi _{2}}=\langle j^{0}({\bf x},t)\rangle
_{{\rm fr}}+\langle j^{0}({\bf x},t)\rangle _{{\rm sc}}+\langle j^{0}({\bf x}%
,t)\rangle _{-}  \label{a6.27}
\end{equation}
where
\begin{equation}
\langle j^{0}({\bf x},t)\rangle _{{\rm fr}}=\frac{e}{(2\pi )^{n}}%
\sum\limits_{\varepsilon _{k^{\prime \prime }},\varepsilon _{k^{\prime
}}=\pm 1}\int \varepsilon _{k^{\prime }}\Psi _{{\rm fr}}^{(+)}({\bf x}%
,t;\varepsilon _{k^{\prime \prime }},P^{\prime \prime })\Psi _{{\rm fr}%
}^{(-)}({\bf x},t;\varepsilon _{k^{\prime }},P^{\prime \prime })dP^{\prime
\prime }+(\hbox{c.c})  \label{a6.28}
\end{equation}
\begin{equation}
\langle j^{0}({\bf x},t)\rangle _{{\rm sc}}=\frac{e}{(2\pi )^{n}}%
\sum\limits_{\varepsilon _{k^{\prime \prime }},\varepsilon _{k^{\prime
}}=\pm 1}\int \varepsilon _{k^{\prime }}\Psi _{{\rm sc}}^{(+)}({\bf x}%
,t;\varepsilon _{k^{\prime \prime }},P^{\prime \prime })\Psi _{{\rm sc}%
}^{(-)}({\bf x},t;\varepsilon _{k^{\prime }},P^{\prime \prime })dP^{\prime
\prime }+(\hbox{c.c})  \label{a6.29}
\end{equation}
\[
\langle j^{0}({\bf x},t)\rangle _{-}=\frac{e}{(2\pi )^{n}}%
\sum\limits_{\varepsilon _{k^{\prime \prime }},\varepsilon _{k^{\prime
}}=\pm 1}\int \Psi _{{\rm fr}}^{(+)}({\bf x},t;\varepsilon _{k^{\prime
\prime }},P^{\prime })(\varepsilon _{k^{\prime }}+\varepsilon _{k^{\prime
\prime }})
\]
\begin{equation}
\Psi _{{\rm sc}}^{(-)}({\bf x},t;\varepsilon _{k^{\prime }},P^{\prime \prime
})dP^{\prime \prime }+(\hbox{c.c})  \label{a6.30}
\end{equation}
Here $\langle j^{0}\rangle _{{\rm fr}}$, $\langle j^{0}\rangle _{{\rm sc}}$
and $\langle j^{0}\rangle _{-}$ are respectively the charge of
noninteracting wave packets, scattering charge and the charge escaping from
the wave packets as a result of the scattering.

Due to the factor $\varepsilon _{k^\prime }$ in Eq. (\ref{a6.29}) the
scattered charge can vanish even in the case, when $\Psi _{sc}$ does not
vanish. But in the case $\varepsilon _{r} =-\varepsilon _{q}$ the charge
\begin{equation}
Q_{-}(t)= \int \langle j^{0}({\bf x},t)\rangle _{-}d{\bf x},  \label{a6.31}
\end{equation}
escaping from each of the wave packets taken separately, vanishes. It means
that in the two-WL case {\it the particle does not interact with antiparticle%
}.

Let us present results of calculations produced by means of the stationary
phase technique.
\begin{equation}
\langle j^{0}({\bf x},t)\rangle _{{\rm fr}}=\frac{e}{(\pi a^{2})^{n/2}}%
\{\varepsilon _{r}\exp [-(x-y-v_{1}t)^{2}/a^{2}]+\varepsilon _{q}\exp
[-(x-z-v_{2}t)^{2}/a^{2}]\}  \label{a6.32}
\end{equation}
\begin{equation}
Q_{{\rm sc}}(t)=-Q_{-}(t)=-\frac{e\lambda \Delta _{1}(R,Q)\varepsilon _{q}}{%
(2\pi a^{2})^{(n-1)/2}E({\bf r})E({\bf q})\sqrt{({\bf v}_{1}-{\bf v}_{1})^{2}%
}|\Delta _{+}(R,Q)|^{2}}  \label{a6.33}
\end{equation}
\noindent where
\begin{equation}
\Delta _{1}(R,Q)=\frac{\Delta _{+}(R,Q)-\Delta _{-}(R,Q)}{2i}  \label{a6.34}
\end{equation}
Calculation of the integral (\ref{a6.2}) leads to the result

\begin{equation}
J=(2\pi a^{2})^{(n-1)/2}\sqrt{1-[({\bf v}_1 ,{\bf v}_2 )^{2}- {\bf v}_1^{2}
{\bf v}_2^{2}]({\bf v}_1 -{\bf v}_2 )^{-2}}  \label{a6.35}
\end{equation}
By means of Eqs. (\ref{a6.1}), (\ref{a6.3}), (\ref{a6.35}) one obtains for
the total scattering section

\begin{equation}
\sigma _{{\rm tot}}=\frac{|Q_{{\rm sc}}|}{J}=\frac{\lambda |\Delta _{1}(R,Q)|%
}{2|\Delta _{+}(R,Q)|\sqrt{(r_{i}q^{i})^{2}-m^{2}}},  \label{a6.36}
\end{equation}
\begin{equation}
r^{i}=\{\varepsilon _{r}E({\bf r}),{\bf r\}},\qquad q^{i}=\{\varepsilon
_{q}E({\bf q}),{\bf q\}}  \label{a6.37}
\end{equation}
Using Eqs.(\ref{a4.49}), (\ref{a4.42}), one obtains in the case $n=2$
\[
\sigma _{{\rm tot}}=\frac{\lambda ^{2}\delta _{\varepsilon _{r},\varepsilon
_{q}}}{8\gamma \sqrt{2(\gamma ^{2}-4m^{2})}[(1-{\frac{\lambda }{8\pi \gamma }%
}\log |\frac{2m+\gamma }{2m-\gamma }|)^{2}+{\frac{\lambda ^{2}}{64\gamma ^{2}%
}}]},
\]
\[
\gamma ^{2}=(r^{i}+q^{i})(r_{i}+q_{i}).
\]
In the case $n=3$ $\Delta _{1}$ is finite and $\Delta _{+}$ diverges. Then $%
\sigma _{{\rm tot}}=0,$ and there is no scattering.

The result (\ref{a6.36}) can be obtained also in the scope of conventional $%
S $-matrix approach. Let us identify the operator $S^{(-)}$ defined by Eq. (%
\ref{a5.38}) with $S$-matrix. The elements of $S$-matrix have the form
\[
_{+}\langle P,K|S^{(-)}|K^{\prime },P^{\prime }\rangle _{+}\equiv
_{+}\langle u,{\bf s|}S^{(-)}|{\bf s^{\prime }},u^{\prime }\rangle _{+}=
\]
\begin{equation}
\delta ({\bf s}-{\bf s^{\prime }})\{{\frac{1}{2}}[\delta (u-u^{\prime
})+\delta (u+u^{\prime })]-2\pi iT({\bf s},u,u^{\prime })\delta \lbrack
\omega _{1}(u,{\bf s})-\omega _{1}(u^{\prime },{\bf s})]\}  \label{a6.38}
\end{equation}
where according to Eqs. (\ref{a5.39}), (\ref{a4.29}), (\ref{a4.36})
\begin{equation}
T({\bf s},u,u^{\prime })=\frac{\lambda }{(2\pi )^{n}}\frac{(\varepsilon
_{k}+\varepsilon _{p})\xi _{2}^{2}(u,{\bf s})}{2\Delta _{-}(w)}
\label{a6.39}
\end{equation}
According to optical theorem [9] (cpt $5,$ f-la $\;(39))$
\begin{equation}
\sigma _{{\rm tot}}=-\frac{2}{v}(2\pi )^{n}{\rm Im}(T({\bf s},u,u))
\label{a6.40}
\end{equation}
where $v$ is the relative velocity of colliding particles. Substituting Eq.(%
\ref{a6.39}) into Eq.(\ref{a6.40}) and using Eq.(\ref{a6.34}), one obtains
the result (\ref{a6.36}).

Thus, without interaction cut-off one has founded a use of the $S$-matrix.
But one should bear in mind that the scattering model obtained is {\it a
result of the quantization model}, {\it where the charge density operator is
chosen in the form }(\ref{a6.9}). If the charge density operator has another
form, then one has, generally, another scattering model even at the same
quantization scheme.

\section{Operators of Physical Quantities}

Among physical quantities
\[
E(t)=\int E(k)b^{\ast }(K,t)b(K,t)dK
\]
\begin{equation}
+{\frac{\lambda }{2(2\pi )^{n}}}\int \int \int \frac{b^{\ast }(K,t)b^{\ast
}(P,t)b(P^{\prime },t)b(K^{\prime },t)}{\sqrt{\beta ({\bf k})\beta ({\bf p}%
)\beta ({\bf k^{\prime }})\beta ({\bf p^{\prime }})}}\delta ({\bf k}+{\bf p}-%
{\bf k^{\prime }}-{\bf p^{\prime }})dKdPdK^{\prime }dP^{\prime }
\label{a7.1}
\end{equation}
\begin{equation}
P_{\alpha }(t)=\int \varepsilon _{k}k_{\alpha }b^{\ast }(K,t)b(K,t)dK,\qquad
\alpha =1,2,3  \label{a7.2}
\end{equation}
\begin{equation}
Q(t)=\int \varepsilon _{k}b^{\ast }(K,t)b(K,t)dK  \label{a7.3}
\end{equation}
only the charge $Q(t)$ does not depend on $t$ due to dynamic equation (\ref
{a4.3}).

Really, calculating $\partial E/\partial t$ and using Eq. (\ref{a4.3}), one
obtains
\begin{eqnarray}
\left. \frac{\partial E}{\partial t}\right| _{t=0} &=&{\frac{i\lambda }{%
2(2\pi )^{n}}}\int \int \int \int {\frac{\varepsilon _{k}E({\bf k}%
)-\varepsilon _{p}E({\bf p})+\varepsilon _{k^{\prime }}E({\bf k^{\prime }}%
)-\varepsilon _{p^{\prime }}E({\bf p^{\prime }})}{\sqrt{\beta ({\bf k})\beta
({\bf p})\beta ({\bf k^{\prime }})\beta ({\bf k^{\prime }})}}}  \nonumber \\
&&\times b^{\ast }(K)b^{\ast }(P)b(K^{\prime })b(P^{\prime })\delta ({\bf k}+%
{\bf p}-{\bf k^{\prime }}-{\bf p^{\prime }})dKdPdK^{\prime }dP^{\prime
}+O_{3,3}.  \label{a7.4}
\end{eqnarray}
$\partial E/\partial t$ vanishes, if operators $b(P^{\prime })$ and $%
b(K^{\prime })$ commute. In the classic case. when $b(K)$ is a $c$-number, $%
\partial E/\partial t=0$ and $E$ conserves. At the commutation relations (%
\ref{a5.36}), (\ref{a5.37}) $\partial E/\partial t\neq 0,$ and $E(t)$ is not
a conservative quantity. The matrix elements of the type $_{+}\langle
P,K|\partial E/\partial t|K^{\prime },P^{\prime }\rangle _{+}$ diverge.

The operator $E(0)$ can be written in the form
\[
E(0)=\int E({\bf k})b_{0}^{\ast }(K)b_{0}(K)dK
\]
\[
+\int \int \{E({\bf k})\mu _{w^{\prime }}^{(-)}(w)+E({\bf k^{\prime }})\mu
_{w}^{(+)}(w^{\prime })+C_{1}(w,w^{\prime })
\]
\[
-\frac{\lambda \eta _{2}(w)\eta _{2}(w^{\prime })}{2(2\pi )^{n}\Delta
_{+}(w)\Delta _{-}(w^{\prime })}\delta ({\bf k}+{\bf p}-{\bf k^{\prime }}-%
{\bf p^{\prime }})\}b_{0}^{\ast }(K^{\prime })b_{0}^{\ast }(P^{\prime
})b_{0}(K)b_{0}(P)dwdw^{\prime }
\]
\[
+\int \int \{E({\bf k})\mu _{S}(w)+C_{2}(w,S)+\frac{\eta _{2}(w)\delta ({\bf %
s}-{\bf k}-{\bf p})}{\Delta _{-}(w)\sqrt{2(2\pi )^{n}\beta ({\bf s})B}}%
\}b_{0}^{\ast }(K)b_{0}^{\ast }(P)b_{1}(S)dwdS
\]
\begin{equation}
+(\hbox{h.c.})+\int \int \{C_{3}(S,S^{\prime })-\frac{\lambda \delta ({\bf s}%
-{\bf s}^{\prime })}{2(2\pi )^{n}\beta _{1}({\bf s})\left| B\right| }%
\}b_{1}^{\ast }(S)b_{1}(S^{\prime })dSdS^{\prime }+O_{3,3}  \label{a7.5}
\end{equation}
where
\[
C_{1}(w,w^{\prime })=\int \mu _{w^{\prime }}^{(-)}(w^{\prime \prime })\mu
_{w}^{(+)}(w^{\prime \prime })E({\bf k^{\prime \prime }})dw^{\prime \prime }
\]
\begin{equation}
C_{2}(w,S)=\int \mu _{w}^{(-)}(w^{\prime \prime })\mu _{S}(w^{\prime \prime
})E({\bf k^{\prime \prime }})dw^{\prime \prime }  \label{a7.6}
\end{equation}
\[
C_{3}(S,S^{\prime })=\int \mu _{S}(w^{\prime \prime })\mu _{S^{\prime
}}(w^{\prime \prime })E({\bf k^{\prime \prime }})dw^{\prime \prime }
\]
In order that $E(t)$ does not depend on $t$, it is sufficient to restrict
the region of integration in the integrals (\ref{a7.6}), adding a factor $%
\delta _{\varepsilon _{k^{\prime \prime }},\varepsilon _{p^{\prime \prime
}}} $ in each integrand. After such correction the expression (\ref{a7.5})
takes the form
\begin{equation}
E(t)=E(0)=\sum_{A=0,1}^{{}}\int E_{A}({\bf k})b_{A}^{\ast
}(K)b_{A}(K)dK+O_{3,3}  \label{a7.7}
\end{equation}
It is worth to note that addition of the factor $\delta _{\varepsilon
_{k^{\prime \prime }},\varepsilon _{p^{\prime \prime }}}$ annihilates matrix
elements between Hilbert spaces ${\cal H}_{2}$ and ${\cal H}_{2}\backslash
{\cal H}_{2,s}$. It prevents from passing from ${\cal H}_{2,s}$ into ${\cal H%
}_{2}$. A like correction should be made in the expansion of $P_{\alpha }(t)$
over operators $b_{A}^{\ast }(K),b_{A}(K^{\prime })$. After this correction $%
P_{\alpha }(t)$ takes the form
\begin{equation}
P_{\alpha }(t)=P_{\alpha }(0)=\int \varepsilon _{k}k_{\alpha }b_{A}^{\ast
}(K)b_{A}(K)dK+O_{3,3},\qquad \alpha =1,2,3  \label{a7.8}
\end{equation}
A like restriction of the integration region should be made in the expansion
of the energy-momentum tensor
\begin{equation}
T^{ik}=\varphi ^{\ast i}\varphi ^{k}+\varphi ^{\ast k}\varphi ^{i}-g^{ik}L
\label{a7.9}
\end{equation}
over operators $b_{A}^{\ast }(K),b_{A}(K^{\prime })$. Then it becomes to
satisfy the conservation law
\begin{equation}
\partial _{k}T^{ik}=O_{3,3}  \label{a7.10}
\end{equation}

\section{ Concluding Remarks}

The problem of pair production is the principal problem of QFT. The
relations obtained in ${\cal H}_{2,s}$ are exact. They are applicable at any
energies of colliding particles. But nonlinear model (\ref{a4.1}), (\ref
{a4.3}) of scalar field does not describe pair production. It does not
describe even particle-antiparticle scattering. This circumstance excites a
feeling of dissatisfaction.

Such a dissatisfaction is based to an extent on the way of thinking
connected with the perturbation theory. In the relativistic QFT the dynamics
of all nonlinear models is investigated by the perturbation theory
technique. Nonlinear term generates a set of diagrams which describe both
creation and annihilation of particles. According to this approach {\it each
nonlinear interaction has to lead to pair production, if} energies of
colliding particles are large enough.

But non-perturbative investigation of the classic description of the pair
production leads to another conclusion. Let us return to the action (\ref
{a1.4}) describing behavior of WL in some external field $f(q)$ which
produces pairs. (For simplicity the gravitational field and electromagnetic
one are absent). The Jacobi-Hamilton equation for the action (\ref{a1.4})
has the form
\begin{equation}
\sqrt{f(q)(\frac{\partial S}{\partial q^{i}}g^{ik}\frac{\partial S}{\partial
q^{k}})}=\alpha b,\qquad b=\hbox{const\qquad }  \label{a8.1}
\end{equation}
Eq.(\ref{a8.1}) at $\alpha >0$ permits spacelike 4-momentum $p_{i}=\partial
S/\partial q^{i}$, provided $f(q)<0$. Hence, it permits the WL turn in time
[5]. In the limit $\alpha \rightarrow 0$ Eq.(\ref{a8.1}) turns to
\begin{equation}
f(q)(\frac{\partial S}{\partial q^{i}}g^{ik}\frac{\partial S}{\partial q^{k}}%
-m^{2}c^{2})=0  \label{a8.2}
\end{equation}
This equation means that the point, where $f(q)=0,$ can be a break point.
Remaining timelike, the WL turns in time direction here, i.e. the external
field $f(q)$ produces or annihilates a pair at this point. In ref.\cite{R70}
example of such solutions is constructed for the finite $\alpha >0.$ At
finite $\alpha >0$ there is a vicinity of the turning point, where WL is
spacelike. At $\alpha \rightarrow +0$ this vicinity degenerates into a break
point, and WL is timelike everywhere except for the break point. At this
point WL changes its time direction.

It was shown also \cite{R70} that the pair production in the external field $%
f(q)$ takes place in the limit $\alpha \rightarrow +0.$ In other words, the
action (\ref{a1.4}) can describe pair production at infinitesimal $\alpha
>0. $

Let us use this circumstance and expand Eq. (\ref{a1.4}) over the parameter $%
\alpha $.
\begin{equation}
S[q]=-\int\limits_{\min (\tau ^{\prime },\tau ^{\prime \prime })}^{\max
(\tau ^{\prime },\tau ^{\prime \prime })}\left\{ mc\sqrt{\dot{q}^{k}g_{kl}%
\dot{q}^{l}}-\frac{\alpha f(q)}{2mc\sqrt{\dot{q}^{k}g_{kl}\dot{q}^{l}}}%
\right\} d\tau  \label{a8.3}
\end{equation}
In this case the Jacobi-Hamilton equation has the form
\begin{equation}
\frac{\partial S}{\partial q^{i}}g^{ik}\frac{\partial S}{\partial q^{k}}=%
\left[ mc+\frac{\alpha b^{2}}{f(q)}\right] ^{2},\qquad b=\hbox{const\qquad }
\label{a8.4}
\end{equation}
Here the 4-momentum $p_{i}=\partial S/\partial q^{i}$ is always timelike.
The WL turn in time is impossible. In the limit $\alpha \rightarrow 0$ the
field $f(q)$ disappears, and the action (\ref{a8.3}) cannot describe pair
production.

The action (\ref{a1.4}) permits spacelike $\dot{q}^{i}$ (at $\alpha f(q)<0)$%
, but the action (\ref{a8.3}) does not permit them in principle. It is a
distinction between the two actions. This example shows that a special
interaction is needed for pair production. It is possible that the
self-action of the model (\ref{a4.1}), (\ref{a4.2}) has not this property.

It seems rather improbable that pair production can appear in the model (\ref
{a4.1}) at consideration of situations in ${\cal H}_{3}$ and ${\cal H}_{4}$
where three and four WLs are considered, because the nonlinear term of the
model associates with the expansion (\ref{a8.3}) over powers of the
interaction constant.

But why does consideration of the model (\ref{a4.1}), using conventional
quantization \cite{GJ68} - \cite{GJ72}, describe pair production? The answer
is very simple. Because the conventional $PA$-scheme of quantization is
inconsistent. When one uses an inconsistent conception, then, exhibiting
enough ingenuity, one may obtain any desired result. At the conventional
quantization according to $PA$-scheme the total world line is separated into
fragments (particles and antiparticles). But one fails to join these
fragments in proper way. Some portion of fragments remains disconnected.
They imitate pair production.

The most unexpected feature of the quantization in terms of WLs is different
simultaneous commutation relations for the free field and for the
self-acting field, (i.e. the simultaneous commutation relations depend on $%
\lambda )$. As a whole this conception {\it does not contain any visible
inconsistencies.}

\end{document}